\documentclass[12pt]{article}
\usepackage{amsmath}
\usepackage{graphicx,psfrag,epsf}
\usepackage{enumerate}
\usepackage{natbib}
\usepackage{url} % not crucial - just used below for the URL 

\newcommand{\blind}{0}

\usepackage{amssymb,amsthm}

\usepackage[svgnames]{xcolor}

\usepackage{soul}
\usepackage{mathrsfs}
\usepackage{stmaryrd}

\usepackage{todonotes}
\usepackage{layout}
\usepackage{mathtools}
\usepackage{wasysym}
\usepackage{paralist}
\usepackage{multirow}
\usepackage{comment}

\usepackage{chngcntr}
\usepackage{apptools}
\AtAppendix{\counterwithin{theorem}{section}}
\AtAppendix{\counterwithin{lemma}{section}}

\usepackage[linesnumbered,ruled]{algorithm2e}
\SetKwInput{KwInput}{Input}                % Set the Input
\SetKwInput{KwOutput}{Output}              % set the Output

\newtheorem{lemma}{Lemma}%[section]

\newtheorem{design}{Design}

\newtheorem{theorem}{Theorem}

\newtheorem{assumption}{Assumption}

\renewcommand{\theassumption}{\Alph{assumption}}
\renewcommand{\thelemma}{\arabic{lemma}}

\usepackage{mathpazo}

\usepackage{xspace}
\usepackage{bm}
\usepackage[nodisplayskipstretch]{setspace}
\usepackage{accents}

\usepackage{footnote}
\usepackage{tabularx}
\usepackage{threeparttable}
\makesavenoteenv{tabular}
\makesavenoteenv{table}

\usepackage{subfig}
\usepackage{microtype}

\usepackage[colorlinks=true, linkcolor=Maroon, urlcolor=Maroon, citecolor=Maroon, hypertexnames=false]{hyperref}

\usepackage{cleveref}
\crefname{figure}{figure}{figures}
\creflabelformat{figure}{#2#1#3}

\crefname{equation}{equation}{equations}
\crefrangeformat{equation}{(#3#1#4) to~(#5#2#6)}
\creflabelformat{equation}{(#2#1#3)}

\crefname{lemma}{lemma}{lemmas}
\creflabelformat{lemma}{#2#1#3}
\crefname{corollary}{corollary}{corollaries}
\creflabelformat{corollary}{#2#1#3}
\crefname{design}{design}{designs}
\creflabelformat{design}{#2#1#3}
\crefname{proposition}{proposition}{propositions}
\creflabelformat{proposition}{#2#1#3}
\crefname{condition}{condition}{conditions}
\creflabelformat{condition}{#2#1#3}
\crefname{assumption}{assumption}{assumptions}
\creflabelformat{assumption}{#2#1#3}
\crefname{remark}{remark}{remarks}
\creflabelformat{remark}{#2#1#3}
\crefname{appendix}{appendix}{appendices}
\creflabelformat{appendix}{#2#1#3}

\renewcommand{\Pr}{\mathbb{P}}

\newcommand{\Exp}{\mathbb{E}}

\newcommand{\BASELINESTRETCH}{1.5}
\newcommand{\one}{\mathbb{1}}

\newcommand\raiseT[2]{\raisebox{0.25ex}{$#1#2$}}
\newcommand\tr{{\mathpalette\raiseT{\intercal}}}

\newcommand{\bl}[1]{\underaccent{\bar}{#1}}

\newcommand{\OR}{\mathrm{OR}}

\newcommand{\sC}{\mathcal{C}}
\newcommand{\CC}{\mathrm{CC}}
\newcommand{\CP}{\mathrm{CP}}
\newcommand{\RR}{\mathrm{RR}}

\newcommand{\AR}{\mathrm{AR}}

\newcommand{\ourtitle}{Causal inference under outcome-based sampling with monotonicity assumptions}

\newcommand{\abstractionthree}{
We study causal inference under case-control and case-population sampling. Specifically, we focus on the binary-outcome and binary-treatment case, where the parameters of interest are causal relative and attributable risks defined via the potential outcome framework. It is shown that strong ignorability is not always as powerful as it is under random sampling and that certain monotonicity assumptions yield comparable results in terms of sharp identified intervals. Specifically, the usual odds ratio is shown to be a sharp identified upper bound on causal relative risk under the monotone treatment response and monotone treatment selection assumptions. We offer algorithms for inference on the causal parameters that are aggregated over the true population distribution of the covariates. We show the usefulness of our approach by studying three empirical examples: the benefit of attending private school for entering a prestigious university in Pakistan; the relationship between staying in school and getting involved with drug-trafficking gangs in Brazil; and the link between physicians’ hours and size of the group practice in the United States.
}

\begin{document}

\bibliographystyle{agsm}

\def\spacingset#1{\renewcommand{\baselinestretch}%
	{#1}\small\normalsize} \spacingset{1}

%%%%%%%%%%%%%%%%%%%%%%%%%%%%%%%%%%%%%%%%%%%%%%%%%%%%%%%%%%%%%%%%%%%%%%%%%%%%%%

\if0\blind
{
	\title{\vspace*{-1cm} \bf \ourtitle}
	\author{Sung Jae Jun\thanks{
			The authors would like to thank the editor, an associate editor, two anonymous referees, Guido Imbens, Chuck Manski, Francesca Molinari and seminar participants at Cemmap, Oxford, and Penn State for helpful comments and Leandro Carvalho and Rodrigo Soares for sharing their dataset and help. This work was supported in part by the European Research Council (ERC-2014-CoG-646917-ROMIA) and by  the UK Economic and Social Research Council (ESRC) through research grant (ES/P008909/1) to the CeMMAP.}\hspace{.2cm}
		\\
		Department of Economics, Pennsylvania State University\\
		and \\
		Sokbae Lee \\
		Department of Economics, Columbia University and IFS}
	\date{October 19, 2023}
	\maketitle
} \fi

\if1\blind
{
	\bigskip
	\bigskip
	\bigskip
	\begin{center}
		{\LARGE\bf \ourtitle}
	\end{center}
	\medskip
} \fi

%\bigskip

%\newpage

\begin{abstract}\noindent
	\abstractionthree
\end{abstract}

\vfill

\newpage
\spacingset{1.8} % DON'T change the spacing!

%%%%%%%%%%%%%%%%%%%%%%%%%%%%%%%%%%%%%%%%%%%%%%%%%%%%%%%%%%%%%%%%%%%%%%%%%%%%%%%%%
%%%%%%%%%%%%%%%%%%%%%%%%%%%%%%%%%%%%%%%%%%%%%%%%%%%%%%%%%%%%%%%%%%%%%%%%%%%%%%%%%%

\section{Introduction}

Random sampling is convenient for causal inference, but it may be too costly in practice for various reasons. For instance, rare events are likely to be severely under-represented in a random sample of a finite size: e.g., cancer \citep{breslow1980statistical}, infant death \citep{currie2005air}, consumer bankruptcy \citep{Domowitz:1999}, entering a highly prestigious university \citep{Delavande:Zafar:19}, and drug trafficking \citep{carvalho2016living}. The objective of this paper is to study causal inference in outcome-based sampling scenarios such as case-control or case-population studies.  

We focus on observational data, as opposed to experimental ones, with a binary outcome and a binary treatment. \cite{Holland:Rubin} adopt the potential outcome framework to show that the assumption of strong ignorability can be used to identify the counterfactual odds ratio in case-control studies. They then argue that the counterfactual odds ratio approximates the ratio of two potential-outcome probabilities (i.e., causal relative risk) under the rare-disease assumption, which says that the probability of outcome occurrence (e.g., having ``a certain disease'') is close to zero. Their work is our starting point, and we make additional contributions in several ways. 

First, we focus on two direct causal parameters (i.e., a ratio or a difference of two potential-outcome probabilities) that are more straightforward to interpret than counterfactual odds ratios: our parameters of interest are the causal relative and attributable risks given a specific value of covariates. Second, we do not appeal to the rare-disease assumption, and we take the perspective of partial identification \citep[see, e.g.,][among others]{manski1995book,manski2003partial,manski2009identification,tamer2010partial,molinari:2020}. Third, we consider a set of monotonicity assumptions, and we compare their identification power with that of strong ignorability. Strong ignorability is a popular setup for causal inference, but its identification power in outcome-based sampling turns out to be somewhat limited.  Specifically, in case-control or case-population studies, strong ignorability is generally not sufficient to point identify the causal relative and attributable risks. We can obtain bounds on them, but they are not much better than those we can obtain in a less restrictive setup using monotonicity. Specifically, we will consider monotone treatment response \citep[][MTR hereafter]{manski1997} and monotone treatment selection \citep[][MTS hereafter]{manski2000monotone}.

Our work builds upon \citet[][Ch 6]{manski2009identification}, who conducts a partial identification analysis for both relative and attributable risks under outcome-based sampling without focusing on causal parameters. The MTR and MTS assumptions as well as other related notions of monotonicity have been extensively used in the literature. For example, see 
\citet{vytlacil2007dummy},
\citet{bhattacharya2008treatment,BSV-2012},
\citet{pearl2009causality},
\citet{vanderweele2009propertise},
\citet{Kreider-et-al:2012},
\citet{jiang2014monotone},
\citet{okumura2014concave},
\citet{choi2017estimation},
\citet{Kim-et-al:2018},
and
\citet{MSV-2019}
among others.

We now discuss the relation of our work with the existing literature on causal inference under outcome-based sampling. 
\citet{maansson2007estimation} point out that the propensity score method has only limited ability to control for confounding factors in case-control studies. Our method does not rely on the propensity score. \citet{rose2011causal} and \citet{van2011targeted} use an assumption that the true case probability is known by a prior study. We focus on the instance of unknown case probability. \citet{CC-handbook-chapter} provide an extensive survey on causal inference in case-control studies, but no discussion on partial identification approaches can be found there. Therefore, possibilities based on partial identification appear to be rather underexplored. \citet{kuroki2010sharp} and \citet{gabriel2020causal} are notable exceptions. \citet{gabriel2020causal} obtain bounds on the causal attributable risk in a variety of scenarios including outcome-dependent sampling with an instrumental variable. But they do not leverage any monotonicity assumption, while we do not consider instrumental variables but we use monotonicity restrictions.  \citet{kuroki2010sharp} is more similar to our work in that they obtain bounds on both the causal relative and attributable risks by using the MTR assumption. Our contributions relative to \citet{kuroki2010sharp} can be highlighted as follows: (1) we exploit not only the MTR but also the MTS assumption, and therefore the bounds are different; (2) we consider case-control sampling as well as case-population sampling; (3) we compare the identification power of the popular assumption of strong ignorability with that of the MTR and MTS assumptions; (4) we consider how to aggregate the causal parameters over the distribution of the covariates; and (5) we provide algorithms for causal inference.

The remaining part of the paper is organized as follows. In \cref{sec:framework} we formally present the setup including the causal parameters of interest and the sampling schemes. \Cref{sec:iden:theory,sec:aggregation,} focus on the causal relative risk and attributable risk to address identification and aggregation. \Cref{sec:causal:inference} covers how to carry out causal inference. \Cref{sec:application} presents three empirical applications. Specifically, by using datasets collected in previous studies, we address new research questions that are not examined in the original papers. All the proofs, discussions on semiparametric efficiency and computational algorithms are in Online Appendix. An accompanying R package is available on the Comprehensive R Archive Network (CRAN) at \href{https://CRAN.R-project.org/package=ciccr}{https://CRAN.R-project.org/package=ciccr}, and the replication files are available at \href{https://github.com/sokbae/replication-JunLee-JBES}{https://github.com/sokbae/replication-JunLee-JBES}.

\section{Preliminaries}\label{sec:framework}
\subsection{Causal parameters}
Let $(Y^*, T^*, X^*)$ be a random vector of a binary outcome, a binary treatment, and covariates of a representative individual. Since we are interested in outcome-based sampling, we assume that a random sample of $(Y^*, T^*, X^*)$ is not available. Instead, we have a sample of $(Y,T,X)$, where the distribution of $(T,X)$ given $Y$ is related with that of $(T^*,X^*)$ given $Y^*$. The exact sampling schemes and related assumptions will be discussed in detail later, and in this subsection, we only focus on the parameters of interest.

For the sake of causal inference, we use the usual potential outcome notation. So, $Y^*(t)$ will be the potential outcome for $T^*=t$, and $Y^*$ can be written as $Y^* = Y^*(1)T^* + Y^*(0)(1-T^*)$. Therefore, our notation extends \citet{chen2001parametric} and \citet{Xie-et-al:JASA} by adding an extra layer of potential outcomes.  The causal effect of the treatment can be measured by either (conditional) relative risk or attributable risk: each of them is defined as follows:
\begin{align}
\theta_\RR(x) 
&:= 
\frac{\Pr\{Y^*(1) = 1 \mid X^*=x\}}{\Pr\{ Y^*(0)=1 \mid X^*= x\}}, \label{eq:CRR} \\
\theta_\AR(x) 
&:= 
\Pr\{Y^*(1)=1 \mid X^*=x\} - \Pr\{Y^*(0)=1\mid X^*=x\}, \label{eq:AR}
\end{align}
provided that the denominator of $\theta_\RR(x)$ is strictly positive. Therefore, $\theta_\AR(x)$ is the usual conditional average treatment effect, whereas $\theta_\RR(x)$ is a causal version of the relative risk parameter.

Relative risk defined by a ratio of ``success'' probabilities has been popular in epidemiology and biostatistics, particularly when the ``success'' is a rare event: if the treatment changes the success probability from $0.01$ to $0.02$, then it is a 100\% increase, though the difference of $0.01$ may suggest an impression that the change was unimportant.  Further, it turns out that $\theta_\RR(x)$ is closely related with the odds ratio (in terms of the observed variables), which has been widely used as a measure of association in case-control studies.

\subsection{Bernoulli sampling}
As we mentioned earlier, we assume that a random sample of $(Y^*, T^*, X^*)$ is unavailable. Instead the researcher has access to a random sample of $(Y,T,X)$, where the distribution of $(Y,T,X)$ is related with that of $(Y^*,T^*,X^*)$ by Bernoulli sampling \citep[e.g.][]{breslow2000semi} that we describe below.

In Bernoulli sampling, the researcher first draws a Beroulli variable $Y$ from a pre-specified marginal distribution, after which she randomly draws $(T,X)$ from some $\mathcal{P}_y$ if and only if $Y=y$; so, $Y$ is an artificial device to decide which subpopulation we will draw $(T,X)$ from. If $\mathcal{P}_y$ is the distribution of $(T^*,X^*)$ conditional on $Y^*=y$, then this is nothing but case-control sampling. Since $h_0 = \Pr(Y=1) \in (0,1)$ is part of the sampling scheme, we will assume that it is known; if not, it can be easily estimated without compromising inferential validity. See online appendices B.1 and B.2 for more details. Before we proceed, we make a common support assumption for simplification. 
\begin{assumption}[Common Support]\label{ass:support}
	The support of $X^*$ and that of $X$ given $Y=y$ for $y=0,1$ coincide; the common support will be denoted by $\mathcal{X}$. 
\end{assumption}

Below we discuss two leading cases of Bernoulli sampling that we focus on throughout the paper. 
\begin{design}[Case-Control Sampling]\label{design1}
	For $y\in\{0,1\}$, $\mathcal{P}_y$ is the distribution of $(T^*,X^*)$ given $Y^*=y$.
\end{design}

\begin{design}[Case-Population Sampling]\label{design2}
	$\mathcal{P}_1$ is the conditional distribution of $(T^*,X^*)$ given $Y^*=1$, whereas $\mathcal{P}_0$ represents the distribution of $(T^*,X^*)$ of the entire population.    
\end{design}

\Cref{design1} is arguably the most popular form of case-control studies \citep[e.g.,][]{breslow1996statistics} and \cref{design2} was referred to as ``contaminated case-control studies'' by \citet{lancaster1996case}: we call the latter design case-population sampling, which is more descriptive.  The case-population sampling design has been used to study drug trafficking \citep{carvalho2016living} and mass demonstrations \citep{rosenfeld_2017} among others.

Note that the distribution of $(T,X)$ is identified from the data, but that of $(T^*,X^*)$ may not. For instance, in \cref{design1}, we have $f_X(x) = f_{X^*|Y^*}(x\mid 1)h_0 + f_{X^*|Y^*}(x\mid 0)(1-h_0) \neq f_{X^*}(x)$, unless $h_0$ is the same as $p_0:=\Pr(Y^*=1)$, i.e., the true probability of the case in the population. Further, $f_{YX}(1,x) = f_{X^*|Y^*}(x\mid 1) h_0 = f_{X^*}(x) \Pr(Y^*=1\mid X^* = x) h_0/p_0$, which yields the likelihood function studied in e.g.\ \citet{Manski:Lerman}. We emphasize that $\Pr(Y=1\mid X=x)$ does not have economic interpretation like $\Pr(Y^* = 1\mid X^*=x)$, where the latter is often specified via domain knowledge in a specific field such as a utility function with an additively separable normal or Gumbel error term.

\section{Identification}\label{sec:iden:theory}

In this section, we study identification of the causal parameters, i.e., $\theta_\RR(x)$ and $\theta_\AR(x)$. Aggregation over $x\in\mathcal{X}$ will be considered later. For the purpose of the identification analysis, we consider two sets of assumptions: one is the standard case of strong ignorability, and the other is an alternative possibility based on monotonicity assumptions. We will see that even strong ignorability is not sufficient to point-identify $\theta_\RR(x)$ or $\theta_\AR(x)$ under case-control sampling, i.e., \cref{design1}. 

We consider the following assumptions.
\begin{assumption}[Overlap]\label{ass:overlap}
	For all $(y,t,s,x) \in \{0,1\}^3 \times \mathcal{X}$, 
	\begin{equation*}
	0 < \Pr\{ Y^*(t) = y, T^* = s \mid X^* = x\} < 1.
	\end{equation*}
\end{assumption}\begin{assumption}[Unconfoundedness]\label{ass:unconfounding}
	For all $(t,x)\in\{0,1\}\times \mathcal{X}$,  
	\begin{equation*}	
	\Pr\{ Y^*(t) = 1\mid T^*=1, X^*=x  \} = \Pr\{ Y^*(t) = 1\mid T^*=0, X^*=x  \}.
	\end{equation*}
\end{assumption}

\Cref{ass:overlap,ass:unconfounding} together constitute strong ignorability, which is a standard setup for causal inference. \Cref{ass:overlap} is stated in terms of the joint probability mass function of $Y^*(t)$ and $T^*$ given $X^*=x$. We do this for a few reasons. First, \cref{ass:overlap} ensures that all the conditional probabilities we consider and their ratios are well-defined: e.g., $\theta_\RR(x)$ is well-defined under \cref{ass:overlap}. Also, it ensures that the distribution of $(Y,T,X)$ has enough overlap to identify $\Pr(T=t\mid Y=y, X=x)$ under each of the two Bernoulli sampling schemes. 

The key component of the strong ignorability setup is \cref{ass:unconfounding}. In the following subsections we will start from clarifying how far \cref{ass:unconfounding} can take us to identify the causal parameters under case-control and case-population sampling.  Although it is standard, strong ignorability does not allow the treatment assignment to be endogenous. Therefore, we consider a set of alternative assumptions under which we study how much we can say about the causal parameters under the two sampling scenarios. 

\begin{assumption}[Monotone Treatment Response]\label{ass:MTR}
	$Y^*(1)\geq Y^*(0)$ almost surely.
\end{assumption}

\begin{assumption}[Monotone Treatment Selection]\label{ass:MTS}
	For all $t\in \{0,1\}$ and $x \in \mathcal{X}$, 
	\begin{equation*}
	\Pr\{ Y^*(t) = 1\mid T^* = 1, X^* = x\} \geq \Pr\{ Y^*(t) = 1 \mid T^* = 0, X^* = x\}.
	\end{equation*}
\end{assumption}

\Cref{ass:MTR} was first proposed by \citet{manski1997}, while \cref{ass:MTS} was used by \citet{manski2000monotone}. \Cref{ass:MTR} says that treatment is potentially beneficial but it never hurts. For instance, if an individual does not earn high income with a college degree, then the person will not be highly paid without a college degree, either. 
%\Cref{ass:MTS} says that, other things being equal, those who have a higher degree are at least as likely to earn high incomes, if their education attainment was randomly assigned, compared to  those who did not have a higher degree.  
\Cref{ass:MTS} states that, all else being equal, individuals with a higher degree are at least as likely to earn high incomes if their educational attainment were randomly assigned, as compared to those without a higher degree.
%So, in substance, the treatment decision chosen by an individual reveals the `type' of the person: i.e.,\SJtodo{changed ';' to ': i.e.,'} those who choose to obtain a higher degree are more motivated, and\SJtodo{a comma added} they would not be any less likely to earn high incomes than those who choose not to obtain a higher degree if they were randomly assigned to a different treatment status.  
In essence, the treatment decision made by an individual reveals their `type': continuing with the same example, those opting for a higher degree are more motivated, and they would be at least as likely to earn high incomes as those who choose not to pursue a higher degree if they were randomly assigned to different educational attainment.
\Cref{ass:MTS} is trivially weaker than \cref{ass:unconfounding}, and it allows individuals with `higher ability' to self-select a higher degree. 

Before we move on, we define the following functions:
\begin{align}
	r_{\CC}(x,p)
	&:=
	\frac{p(1-h_0)\Pr(Y=1\mid X=x)}{p(1-h_0)\Pr(Y=1\mid X=x) + h_0(1-p)\Pr(Y=0\mid X=x)}, \label{def rcc}\\
	r_\CP(x,p) 
	&:=
	\frac{p(1-h_0)}{h_0}\frac{\Pr(Y=1\mid X=x)}{\Pr(Y=0\mid X=x)}, \label{def rcp}
\end{align}
where $\Pr(Y=1\mid X=x)$ is the prospective regression function identified from the data. Here, both $r_\CC(x,p)$ and $r_\CP(x,p)$ can be alternatively expressed by using the conditional densities of $X$ given $Y=y$ by the Bayes rule, which is related with the distribution of $X^*$ given $Y^*=y$ or simply the distribution of $X^*$, depending on the sampling design. Indeed, it can be shown that $r_\CC(x,p_0) = \Pr(Y^*=1\mid X^*=x)$ under case-control sampling and $r_\CP(x,p_0) = \Pr(Y^*=1\mid X^*=x)$ under case-population sampling, where $p_0 = \Pr(Y^*=1)$: see lemma A.3 in the online appendix. Therefore, one can view the functions $r_\CC$ and $r_\CP$ as devices to exploit the fact that the only unidentified object in our context will be $p_0$.

\subsection{Causal relative risk}\label{sec:id}
In this section, we study identification of $\theta_\RR(x)$, for which we first introduce some notation. Let $\Pi(t\mid y,x) = \Pr(T=t \mid Y=y, X=x)$ be the retrospective regression function. For $(x,p)\in \mathcal{X}\times [0,1]$ and for $d\in \{\CC,\CP\}$, define
\[ 
\Gamma_{d,\RR}(x,p)
:=
\frac{\Pi(1\mid 1,x)}{\Pi(0\mid 1,x)}\times 
\frac{ \Pi(0\mid 0,x) + r_d(x,p) \{\Pi(0\mid 1,x) - \Pi(0\mid 0, x)\}   }{ \Pi(1\mid 0,x) + r_d(x,p) \{\Pi(1\mid 1,x) - \Pi(1\mid 0, x)\} },
\] 
where \cref{ass:overlap} ensures that $\Pi(t\mid y,x) \neq 0$ for all $(t,y,x)\in \{0,1\}\times \{0,1\}\times \mathcal{X}$ in each of the two Bernoulli sampling schemes. It is worth noting that $\Gamma_{d,\RR}(x,0)$ for both $d\in\{\CC,\CP\}$ is just the covariate-adjusted odds ratio, i.e.,
\[
\OR(x) 
:= 
\frac{\Pi(1\mid 1,x)}{\Pi(0\mid 1,x)}\frac{\Pi(0\mid 0,x)}{\Pi(1\mid 0,x)}
,
\]
which is a popular measure of covariate-adjusted association in case-control studies. Since $\OR(x)$ is more descriptive than $\Gamma_{d,\RR}(x,0)$, we will use the former notation whenever it is relevant.

The following lemma shows what we could achieve if we had a random sample, i.e., if $(Y^*,T^*,X^*)$ were observed.
\begin{lemma}[RR-Benchmark]\label{lem:rr benchmark}
If \cref{ass:support,ass:overlap,ass:unconfounding} are satisfied, then for all $x\in \mathcal{X}$,
\[
\theta_\RR(x) = \frac{\Pr(Y^*=1\mid T^*=1, X^*=x)}{\Pr(Y^*=1\mid T^*=0, X^*=x)}.
\]
Alternatively, if \cref{ass:support,ass:overlap,ass:MTR,ass:MTS} are satisfied, then for all $x\in\mathcal{X}$,
\[
1\leq \theta_\RR(x) \leq \frac{\Pr(Y^*=1\mid T^*=1, X^*=x)}{\Pr(Y^*=1\mid T^*=0, X^*=x)},	
\] 
where the bounds are sharp. 
\end{lemma}
\Cref{lem:rr benchmark} serves two purposes. First, it is useful as a middle step to establish the sharp identifiable bounds under Bernoulli sampling, i.e., under designs \ref{design1} (Case-Control) and \ref{design2} (Case-Population). Second, it shows benchmark results for the identification of $\theta_\RR(x)$ in that it shows the best we can achieve under random sampling through unconfoundedness or monotonicity. Therefore, \cref{lem:rr benchmark} should be compared with \cref{thm:unconfounding,cor:unconfounding,thm:bounds} that are discussed below.

Point identification under random sampling and strong ignorability is not surprising. Partial identification under random sampling and the monotonicity assumptions is reminiscent of e.g.\ \citet{manski2000monotone}. However, in our setup, the researcher does not have access to a random sample of $(Y^*,T^*,X^*)$, and therefore, \cref{lem:rr benchmark} is not an identification result. It will serve as a benchmark to show the cost of case-control or case-population studies in terms of identification.

Recall that $p_0 = \Pr(Y^*=1)$ is the true probability of the case, which is an unidentified object under Bernoulli sampling. 
\begin{theorem}\label{thm:unconfounding}
Suppose that \cref{ass:support,ass:overlap,ass:unconfounding} are satisfied. Then, for all $x\in \mathcal{X}$, we have the following.
\begin{enumerate}[(1)]
\item Under case-control sampling, i.e., \cref{design1}, we have $\theta_\RR(x) = \Gamma_{\CC,\RR}(x, p_0)$.
\item Under case-population sampling, i.e., \cref{design2}, we have $\theta_\RR(x) = \OR(x)$. 
\end{enumerate}
\end{theorem}

\Cref{thm:unconfounding} is not identification results in the case of case-control sampling: $p_0$ is unidentified in \cref{design1}. In contrast, it shows that $\theta_\RR(x)$ is point identified under case-population sampling. Therefore, \cref{design2} provides an easier environment for causal inference, at least under unconfoundedness.  It seems ironic that \cref{design2} was referred to as case-control sampling with contamination by \citet{lancaster1996case} but that the `contamination' is in fact helpful for identification.

In the case of \cref{design1} we do not have point identification, but there is only one simple parameter that is unidentified. Therefore, it is not too difficult to proceed with a partial identification approach. We will further elaborate about this possibility. Before we proceed though, it is worth comparing the case-control case of \cref{thm:unconfounding} with \citet{Holland:Rubin}. Specifically, \citet{Holland:Rubin} show that under \cref{design1}, $\OR(x)$ is equal to the odds ratio in terms of the potential outcomes if strong ignorability is imposed: i.e.,
\begin{equation}\label{eq:holland-rubin}
	\OR(x)
	=
	\frac{\Pr\{ Y^*(1) = 1 \mid X^*=x \}}{\Pr\{ Y^*(0) = 1 \mid X^*=x \}}
	\frac{\Pr\{ Y^*(0) = 0 \mid X^*=x \}}{\Pr\{ Y^*(1) = 0 \mid X^*=x \}}.	
\end{equation}
\Cref{eq:holland-rubin} is an identification result, but its right-hand side expression is not straightforward to interpret. 
%It seems that the reason why \citet{Holland:Rubin} focused on the right-hand side expression of \cref{eq:holland-rubin}, rather than the causal relative risk $\theta_\RR(x)$ that is easier to interpret, is that the former is identified by $\OR(x)$ while the latter requires that we deal with the fact that $p_0$ is unidentified.
It appears that the reason \citet{Holland:Rubin} emphasized the right-hand side expression of \cref{eq:holland-rubin} instead of the more easily interpretable causal relative risk $\theta_\RR(x)$ is that the former is identified by $\OR(x)$, whereas the latter necessitates addressing the issue that $p_0$ remains unidentified.

Generally, $\Gamma_{\CC,\RR}(x,p_0)$ is different from $\OR(x) = \Gamma_{\CC,\RR}(x,0)$. However, this issue has been traditionally ignored, because if $Y^*$ represents a rare event in that $p_0\approx 0$, then $\Gamma_{\CC,\RR}(x,p_0) \approx \Gamma_{\CC,\RR}(x,0)$ by continuity: the assumption of small $p_0$ is known as the rare disease assumption in epidemiology. However, the quality of the approximation via continuity can quickly decrease as $p_0$ deviates from zero, i.e., the occurrence of $Y^*=1$ becomes less uncommon in the population. Therefore, when $p_0$ is away from zero, a natural alternative approach is to take a partial identification approach, where we target the function $\Gamma_{\CC,\RR}(x,\cdot)$ itself, at least within a certain neighborhood of $0$. 

Below we will write $f_{A,B}(a,b)$ for the Radon-Nikodym density of $A,B$ (with respect to some dominating measure). For instance, when $A$ is discrete and $B$ is continuous, we will have $f_{A,B}(a,b) = \Pr(A=a) f_{B|A}(b\mid a)$ by using a product of count and Lebesgue measures. Similarly, $f_{A,B|C}(a,b\mid c)$ will be used to denote a conditional density of $(A,B)$ at $(a,b)$ given $C=c$.

\begin{assumption}\label{ass:rare}
	There is a known value $\bar p$ such that $p_0 \leq \bar p$, where $\bar p \leq 1$ under \cref{design1}, and $\bar p \leq \bar p^*$ with
	\begin{equation} \label{eq:pbarstar}
		\bar p^* 
		:= 
		\inf\ \Biggl\{ \frac{f_{T,X|Y}(t,x\mid 0)}{f_{T,X|Y}(t,x\mid 1)}:\ \text{$t,x$ are such that $f_{T,X|Y}(t,x\mid 1)>0$} \Biggr\}
	\end{equation}
	under \cref{design2}.   
\end{assumption}
We remark that $\bar p^*\leq 1$: see online appendix F. Under case-control or case-population sampling, $p_0 = \Pr(Y^*=1)$ is generally unidentified, because $Y^*$ is not randomly observed. Since case-control or case-population sampling is popular when $Y^*=1$ is a rare event and therefore a random sample of a modest size tends to contain too few observations of the case of interest, we do not want to rule out the possibility that $p_0$ is close to zero: it is straightforward though to replace \cref{ass:rare} with the one that $p_0\in [\bl{p}, \bar p]$ for some known values of $\bl{p}$ and $\bar p$.

If we have an auxiliary sample, from which we learn about $p_0$, then plugging that piece of information into the case-control sample will resolve the identification problem since $p_0$ is the only unidentified object here. Even if it is difficult to pin down $p_0$ exactly, we may have external sources or qualitative information about how prevalent a certain `disease' is, and such information can be used to place an upper bound on $p_0$. Relying on the researcher's prior knowledge on an unidentified object has been used in the context of robust estimation as well \citep[e.g.,][]{horowitz1995identification,horowitz199716}. 

Choosing $\bar p = 1$ in \cref{design1} corresponds to the case where the researcher has no prior information for $p_0$ at all: we do not rule out this possibility. In \cref{design2}, it may be possible to find $\bar p<1$ even without having any external source of information at all. To see this point, we note that under \cref{design2}, we must have 
\begin{align*}
        f_{T,X|Y}(t,x\mid 0)
        =
        f_{T^*,X^*}(t,x)
        &=
        f_{T^*,X^*|Y^*}(t,x\mid 1)p_0 +     f_{T^*,X^*|Y^*}(t,x\mid 0)(1-p_0) \\
        &=
        f_{T,X|Y}(t,x\mid 1)p_0  +   f_{T^*,X^*|Y^*}(t,x\mid 0)(1-p_0),    
\end{align*}
where $f_{T^*,X^*|Y^*}(t,x\mid 0)(1-p_0) = f_{T,X|Y}(t,x\mid 0) - f_{T,X|Y}(t,x\mid 1)p_0 \geq 0$ for all $t,x$. This motivates the definition of $\bar p^*$ in \cref{eq:pbarstar}.

\begin{theorem}\label{cor:unconfounding}
	Suppose that \cref{ass:support,ass:overlap,ass:unconfounding,ass:rare} are satisfied. Under case-control sampling, i.e., \cref{design1}, we have
	\[
	\min\{ \OR(x),\ \Gamma_{\CC,\RR}(x,\bar p)\} 
	\leq 
	\theta_\RR(x) 
	\leq 
	\max\{ \OR(x),\ \Gamma_{\CC,\RR}(x,\bar p)\},
	\]
	and the bounds are sharp. 
\end{theorem}

\Cref{cor:unconfounding} is a simple corollary from \cref{thm:unconfounding}, where it is addressed that $p_0$ is unidentified under case-control sampling. Since $\Gamma_{\CC,\RR}(x,p)$ is monotonic in $p\in [ 0, \bar p ]$, it suffices to consider the two end points to obtain sharp bounds, where one of the end points is the odds ratio $\OR(x) = \Gamma_{\CC,\RR}(x,0)$. We also remark that it can be verified that  $\Gamma_{\CC,\RR}(x,\bar p) \geq 0$ because $0\leq r_\CC(x,\bar p)\leq 1$ by definition: this should not be surprising because $\theta_\RR(x) \geq 0$ by definition. 

If \cref{ass:MTR}  is satisfied in addition, then we can show that $\Gamma_{\CC,\RR}(x,\cdot)$ is a decreasing function and therefore it follows that $\Gamma_{\CC,\RR}(x,\bar p) \leq \theta_\RR(x) \leq \Gamma_{\CC,\RR}(x,0) = \OR(x)$ under \cref{design1}. Therefore, the odds ratio represents the maximum causal relative risk that is consistent with what is observed in a case-control study. If there is no information for $p_0$ at all, then the lower bound is simply one. Below we will see that the sharp identifiable bounds $[1,\ \OR(x)]$ on $\theta_\RR(x)$ can still be obtained without relying on the ignorability assumptions in case-control studies.

Unconfoundedness is a popular assumption for causal inference, but it is not always satisfied in observational studies. Further, unlike the standard case of random sampling, it does not deliver point-identification under case-control studies. \Cref{ass:MTR,ass:MTS} provide an alternative possibility, where we do not lose much in terms of partial identification.

\begin{theorem}\label{thm:bounds}
	Suppose that \cref{ass:support,ass:overlap,ass:MTR,ass:MTS} are satisfied. Then, under both \cref{design1,design2}, we have $1\leq \theta_\RR(x) \leq \OR(x)$, where the bounds are sharp. 
\end{theorem}

Unlike \cref{thm:unconfounding,cor:unconfounding}, \cref{thm:bounds} considers the case where we do not have unconfoundedness but we only impose monotonicity. Now, $\OR(x)$ is a sharp upper bound on $\theta_\RR(x)$ under both case-control and case-population sampling designs. 

It is not explicit in \cref{thm:bounds}, but its proof shows that the knowledge of $p_0$ is potentially useful in \cref{design1} but not in \cref{design2}. In fact, if $p_0$ were known, then the sharp bounds on $\theta_\RR(x)$ under \cref{design1} would be given by $[ 1,\ \Gamma_{\CC,\RR}(x,p_0) ]$, whereas those under \cref{design2} would still be $[1,\ \OR(x)]$. This difference arises because a few applications of the Bayes rule show that the sharp upper bound under random sampling, i.e., the prospective regression ratio $\Pr(Y^*=1\mid T^*=1,X^*=x)/\Pr(Y^*=1\mid T^*=0,X^*=x)$ in \cref{lem:rr benchmark}, is equal to $\Gamma_{\CC,\RR}(x,p_0)$ under \cref{design1}, whereas it is equal to $\Gamma_{\CP,\RR}(x,0) = \OR(x)$ under \cref{design2}. Therefore, if we do not have a random sample, but we have access only to a case-control sample, then there is an information loss in terms of sharp identifiable bounds on $\theta_\RR(x)$. In contrast, a case-population sample is equally informative for $\theta_\RR(x)$ as a random sample.  Thus, \cref{design2} provides a better environment for causal inference than \cref{design1} under monotonicity, similarly to the case of unconfoundedness: see our comments below \cref{thm:unconfounding}. The extra challenge in case-control studies can be addressed by the fact that $\Gamma_{\CC,\RR}(x,p)$ is decreasing in $p$. Therefore, the sharp upper bound on $\theta_\RR(x)$ under \cref{design1} is given by the maximum (over $p$) of $\Gamma_{\CC,\RR}(x,p)$, which is equal to $\Gamma_{\CC,\RR}(x,0)$ even without using \cref{ass:rare}.

We now compare \Cref{thm:bounds} with \Cref{thm:unconfounding,cor:unconfounding}. The identification power of strong ignorability depends on the specific sampling design, whereas that of the monotonicity assumptions is independent of which of the two sampling scenarios applies.  Specifically, in case-population studies, i.e., \cref{design2}, unconfoundedness is informative in that it ensures that $\theta_\RR(x)$ is point identified by the odds ratio. However, in case-control studies, i.e., \cref{design1}, unconfoundedness only yields interval identification, where the sharp identifiable bounds are the same as what the monotonicity assumptions can deliver if we have no information for $p_0$.

\subsection{Causal attributable risk}\label{appx:ar}

We now turn to the alternative causal parameter $\theta_\AR(x)$. We need some extra notation. For $(x,p) \in \mathcal{X}\times [0,1]$ and for $d\in \{\CC,\CP\}$, define 
\[ 
	\Gamma_{d,\AR}(x,p)
	:=
	\sum_{j=0}^1
	\frac{(-1)^{j+1}\Pi(j\mid 1,x)}{\Pi(j\mid 0,x) + r_d(x,p)\{\Pi(j\mid 1,x) - \Pi(j\mid 0, x)\}},	
\]
where $\Pi(t\mid y,x)$ and $r_d(x,p)$ are defined in the beginning of \cref{sec:id}. Note that $\Gamma_{d,\AR}(x,0)$ is not exactly the odds difference, though it is similar: it is a difference between two ratios of retrospective regressions.  

We start with the benchmark case of what if we could observe $(Y^*,T^*,X^*)$.

\begin{lemma}[AR-Benchmark]\label{lem:ar benchmark}
If \cref{ass:support,ass:unconfounding} are satisfied, then for all $x\in \mathcal{X}$,
\[
\theta_\AR(x) = \Pr(Y^*=1\mid T^*=1, X^*=x) - \Pr(Y^*=1\mid T^*=0, X^*=x). 	
\]
Alternatively, if \cref{ass:support,ass:overlap,ass:MTR,ass:MTS} are satisfied, then for all $x\in \mathcal{X}$,
\[
0\leq \theta_\AR(x) \leq \Pr(Y^*=1\mid T^*=1, X^*=x) - \Pr(Y^*=1\mid T^*=0, X^*=x),	
\]
where the bounds are sharp. 
\end{lemma}

Similarly to \cref{lem:rr benchmark}, \cref{lem:ar benchmark} has two purposes. First, it is a middle-step result to establish the sharp identifiable bounds on $\theta_\AR(x)$ when we do not have a random sample but only a sample from either \cref{design1} or \cref{design2} is available. Second, it shows benchmark results for the identification of $\theta_\AR(x)$ via unconfoundedness or monotonicity under random sampling. Point identification of $\theta_\AR(x)$ via strong ignorability under random sampling is now a standard result. If strong ignorability is replaced with the monotonicity assumptions, then the regression difference should be interpreted as a sharp upper bound on the causal attributable risk. Below we extend these results to the cases of case-control and case-population sampling.

\begin{theorem}\label{thm:unconfounding AR}
Suppose that \cref{ass:support,ass:overlap,ass:unconfounding} are satisfied. Then, for all $x\in \mathcal{X}$, we have the following.
\begin{enumerate}[(1)]
\item Under case-control sampling, i.e., \cref{design1}, we have $\theta_\AR(x) = r_\CC(x,p_0) \Gamma_{\CC,\AR}(x,p_0)$.
\item Under case-population sampling, i.e., \cref{design2}, we have $\theta_\AR(x) = r_\CP(x,p_0) \Gamma_{\CP,\AR}(x,0)$. 
\end{enumerate}
\end{theorem}

Unlike the case of $\theta_\RR(x)$, $\theta_\AR(x)$ remains unidentified even in \cref{design2}. This happens because $\Gamma_{\CP,RR}(x,0)$ is a ratio of two terms, where $r_\CP(x,p_0)$ cancels out, but $\Gamma_{\CP,\AR}(x,0)$ is a difference and the common factor $r_\CP(x,p_0)$ does not disappear.  Also, unlike $\theta_\RR(x)$, the rare disease approximation does not provide anything useful in either of the two sampling schemes: if $p_0\approx 0$, then $r_\CC(x,p_0) \approx 0$ and $r_\CP(x,p_0) \approx 0$ by continuity. However, the partial identification approach still remains useful.

\begin{theorem}\label{cor:unconfounding AR}
	Suppose that \cref{ass:support,ass:overlap,ass:unconfounding,ass:rare} are satisfied. For all $x\in \mathcal{X}$, we have the following. 
 	\begin{enumerate}[(1)]
 	\item Under case-control sampling, i.e., \cref{design1},
 	\[
 	\min_{p\in [0,\bar p]} r_\CC(x,p) \Gamma_{\CC,\AR}(x,p)\leq \theta_\AR(x)\leq \max_{p\in[0, \bar p]} r_\CC(x,p)\Gamma_{\CC,\AR}(x,p),
 	\]
 	where the bounds are sharp.
 	\item Under case-population sampling, i.e., \cref{design2},
 	\[
		\min\bigl\{ 0,\ r_\CP(x,\bar p)\Gamma_{\CP,\AR}(x,0)\bigr\}
        \leq 
        \theta_\AR(x)
        \leq 
        \max\bigl\{ 0,\ r_\CP(x,\bar p)\Gamma_{\CP,\AR}(x,0)\bigr\},
 	\]	
 	where the bounds are sharp. 
	\end{enumerate}
\end{theorem}

Since $\theta_\AR(x)$ is a difference of probabilities, it is always between $-1$ and $1$. Indeed, we show in the proof that all the bounds in \cref{cor:unconfounding AR} lie within the interval between $-1$ and $1$. \Cref{cor:unconfounding AR} is a simple corollary of \cref{thm:unconfounding AR}: sharpness follows from the fact that $p_0$ is unidentified and that $r_\CC(x,p)\Gamma_{\CC,\AR}(x,p)$ and $r_\CP(x,p) \Gamma_{\CP,\AR}(x,p)$ are all continuous in $p$.  Unlike the case of random sampling, the conditional average treatment effect is only partially identified even under strong ignorability. Also, it is noteworthy that in \cref{design2}, the sign of $\theta_\AR(x)$ is determined by that of $\Gamma_{\CP,\AR}(x,0)$: if we know that $\theta_\AR(x)\geq0$, then we know that the conditional average treatment effect is at most $r_\CP(x,\bar p) \Gamma_{\CP,\AR}(x,0)$.  

We now consider replacing unconfoundedness with the monotonicity assumptions.

\begin{theorem}\label{thm:bounds AR}
	Suppose that \cref{ass:support,ass:overlap,ass:MTR,ass:MTS,ass:rare} are satisfied. Then, for all $x\in \mathcal{X}$, we have the following.
	\begin{enumerate}[(1)]
		\item Under case-control sampling, i.e., \cref{design1}, 
		\[ 
			0 
			\leq \theta_\AR(x) 
			\leq \max_{p\in[0,\bar p]}  r_\CC(x,p) \Gamma_{\CC,\AR}(x,p),
		\] 
		where the bounds are sharp.
		\item Under case-population sampling, i.e., \cref{design2}, 
		\[ 
			0
			\leq 
			\theta_\AR(x) 
			\leq r_\CP(x, \bar p) \Gamma_{\CP,\AR}(x,0),
		\]
		the bounds are sharp.
	\end{enumerate}
\end{theorem}

Similarly to our comments below \cref{thm:bounds}, knowledge of $p_0$ is potentially useful to improve the bounds given in \cref{thm:bounds AR}: this point will be relevant when we discuss aggregation in the following section. This is so because, by the Bayes rule, the difference between the two prospective regression functions that appear in \cref{lem:ar benchmark} can be shown to be equal to $r_\CC(x,p_0)\Gamma_{\CC,\AR}(x,p_0)$ under \cref{design1} and to $r_\CP(x,p_0)\Gamma_{\CP,\AR}(x,0)$ under \cref{design2}, respectively. However, $p_0$ is unrestricted in general, and hence maximizing over $p_0\in [0, \bar p]$ under \cref{ass:rare} delivers the sharp upper bounds.

The bounds in \cref{thm:bounds AR} are comparable with those in \cref{cor:unconfounding AR}. In case-control or case-population sampling, strong ignorability is not as powerful as in random sampling. First, strong ignorability does not deliver point identification of the conditional average treatment effect. Second, the monotonicity assumptions do restrict the sign of $\theta_\AR(x)$, but, otherwise, they have the same amount of information as the strong ignorability assumptions in terms of the maximum admissible value of $\theta_\AR(x)$.

\section{Aggregation}\label{sec:aggregation}

Conditioning on a specific value of the covariate vector and aiming at $\theta_\RR(x)$ or $\theta_\AR(x)$ as in \cref{cor:unconfounding,cor:unconfounding AR,thm:bounds,thm:bounds AR} is one natural approach to deal with potential heterogeneity in the causal treatment effect. However, the corresponding bounds as functions of $x$ (e.g., $\OR(x)$) are complicated objects, and they are difficult to estimate with high precision when $X^*$ is multi-dimensional. 

To avoid the curse of dimensionality, it is popular in case-control studies to adopt logistic regression. Some authors have alternatively parametrized the odds ratio function itself in case-control studies, focusing on establishing a doubly robust estimator of the odds ratio: see e.g.\ \citet{yun2007semiparametric} and \citet{tchetgen2013closed}. Direct parametrization of $\Gamma_{d,\AR}(x,p)$ appears to be uncommon though. 

Parametric assumptions are convenient, but they are restrictive: e.g.\ $\OR(x)$ is generally an unknown function of  $x$ that can be highly nonlinear.  Instead of introducing any parametrization, aggregation over the population distribution of the covariates can be a useful approach to obtain a robust summary measure. 

If one wants to report an aggregated parameter such as $\int_\mathcal{X} \theta_\AR(x) \omega(x) dx$ for some weight function $\omega$, sharp bounds can be obtained by taking max/min over $p_0$ \emph{after} aggregation.  The most natual choice of the weight function $\omega$ is probably the true population density of $X^*$.  The distribution of $X^*$ is unidentified in case-control studies, but the situation is not too bad because the only unidentified object is, again, $p_0$.

Consider the following aggregated parameters:
\begin{equation} \label{eq:aggregated parameters}
    \bar \vartheta_\RR := \int_{\mathcal{X}} \log\bigl\{ \theta_\RR(x) \bigr\} f_{X^*}(x)\ dx
    \quad\text{and}\quad
    \bar \vartheta_\AR = \int_{\mathcal{X}} \theta_\AR(x) f_{X^*}(x)\ dx.
\end{equation}
$\bar \vartheta_\AR$ is the standard average treatment effect. For $\bar \vartheta_\RR$, we use the logarithm of $\theta_\RR(X^*)$ to take an average. Since $\Exp\{ \log\OR(X^*)\} \leq \log\Exp\{ \OR(X^*)\}$ by Jensen's inequality, the average of the logarithm is less likely to be affected unduly by outliers. We also note that it is more conventional to work with the logarithm of the odds ratio than the odds ratio itself.  If one still prefers aggregating $\theta_\RR(x)$ itself, it is straightforward to modify our methodology by using the same principle outlined in this section.

Our approach is to use the fact that the only missing piece in case-control or case-population samples is $p_0$. We first derive sharp identifiable bounds on $\theta_\RR(x)$ and $\theta_\AR(x)$ with $p_0$ given. We then aggregate over the distribution of $X^*$, which depends on $p_0$ in case-control studies. Specifically, we use the fact that for all $x\in \mathcal{X}$,
\[
f_{X^*}(x)
=
\left\{
    \begin{aligned}
        &f_{X|Y}(x\mid 1) p_0 + f_{X|Y}(x\mid 0) (1-p_0) \quad &&\text{in case-control studies,}\\
        &f_{X|Y}(x\mid 0) \quad &&\text{in case-population studies.}
    \end{aligned}
\right.    
\]
We can then rely on \cref{ass:rare} to address the fact that $p_0$ is unidentified. For this purpose, we can maximize or minimize over $p_0\in [0,\bar p]$ to obtain bounds, or, more informatively, we can plot the whole bound functions on $[0, \bar p]$: choosing the maximal value that is allowed for $\bar p$ (e.g., $\bar p = 1$ in case-control studies) corresponds to the case where we have no information for $p_0$. This line of reasoning leads to the main results in this section.  

We will use the following objects: for $d\in \{\CC,\CP\}$, 
\[ 
	\Psi_{d,\RR}(p,y) := \Exp\{ \log\Gamma_{d,\RR}(X,p) \mid Y=y \}.
\] 
The logarithm in the definition of $\Psi_{d,\RR}(p,y)$ is because $\bar\vartheta_\RR$ is the aggregation of $\log\theta_\RR(x)$. If one wants to bound $\int_\mathcal{X} \theta_\RR(x) f_{X^*}(x) dx$, then changing the definition of $\Psi_{d,\RR}(p,y)$ to $\Exp\{ \Gamma_{d,\RR}(X,p) \mid Y=y \}$ will do. Also, we note that $\int_\mathcal{X} \theta_\RR(x) f_{X^*}(x) dx$ differs from the ratio of unconditional counterfactual probabilities.  Let 
\[
\begin{aligned}
	\Psi_{\CC,\AR}(p,y) &:= \Exp\{ r_\CC(X,p)\Gamma_{\CC,\AR}(X,p) \mid Y = y \},\\
	\Psi_{\CP,\AR}(p) &:= \Exp\{ r_\CP(X,p) \Gamma_{\CP,\AR}(X,0) \mid Y=0 \bigr\},
\end{aligned}
\]
where we note that $\Psi_{\CP,\AR}(p)$ is a simple linear function of $p$ by definition. Finally, for $k\in \{\RR,\AR\}$, define $\sC_{\CC,k}(p)$ by a convex combination of $\Psi_{\CC,k}(p,1)$ and $\Psi_{\CC,k}(p,0)$: i.e., $\sC_{\CC,k}(p) := \Psi_{\CC,k}(p,1)p + \Psi_{\CC,k}(p,0)(1-p)$.

\begin{theorem}\label{cor:aggregation1}
Suppose that \cref{ass:support,ass:overlap,ass:unconfounding,ass:rare} are satisfied. We then have the following.
\begin{enumerate}[(1)]
\item Under case-control sampling, i.e., \cref{design1}, the sharp identified bounds on $\bar\vartheta_\RR$ and $\bar\vartheta_\AR$ are given by  
\begin{align*} 
	\min_{p\in[0,\bar p]} \sC_{\CC,\RR}(p) &\leq \bar \vartheta_\RR \leq \max_{p\in[0,\bar p]} \sC_{\CC,\RR}(p),\\
	\min_{p\in[0,\bar p]} \sC_{\CC,\AR}(p) &\leq \bar \vartheta_\AR \leq \max_{p\in[0,\bar p]} \sC_{\CC,\AR}(p).
\end{align*} 

\item Under case-population sampling, i.e., \cref{design2}, we have $\bar \vartheta_\RR = \Psi_{\CP,\RR}(0,0)$, where we remark that this point identification result does not require \cref{ass:rare}. Further, the sharp identified bounds on $\bar\vartheta_\AR$ are given by 
\[ 
	\min\left\{ 0,\ \Psi_{\CP,\AR}(\bar p) \right\} 
	\leq  
	\bar\vartheta_\AR 
	\leq 
	\max\left\{ 0,\ \Psi_{\CP,\AR}(\bar p) \right\}.
\] 
\end{enumerate} 
\end{theorem}

\begin{theorem}\label{cor:aggregation2}
Suppose that \cref{ass:support,ass:overlap,ass:MTR,ass:MTS,ass:rare} are satisfied. Then, we have the following. 
\begin{enumerate}[(1)] %\setcounter{enumi}{2}
\item Under the case-control sampling, i.e., \cref{design1}, the sharp identified bounds on $\bar\vartheta_\RR$ and $\bar\vartheta_\AR$ are given by 
\begin{equation*} 
	0\leq \bar\vartheta_\RR \leq \max_{p\in [0,\bar p]} \sC_{\CC,\RR}(p)
    \quad \text{and} \quad
	0\leq \bar\vartheta_\AR \leq \max_{p\in [0,\bar p]} \sC_{\CC,\AR}(p).
\end{equation*}  

\item Under the case-population sampling, i.e., \cref{design2}, the sharp identified bounds on $\bar\vartheta_\RR$ and $\bar\vartheta_\AR$ are given by 
\begin{equation*} 
0 \leq \bar\vartheta_\RR \leq \Psi_{\CP,\RR}(0,0)
\quad \text{and} \quad 
0 \leq \bar\vartheta_\AR \leq \Psi_{\CP,\AR}(\bar p),
\end{equation*} 
where we remark that the bounds on $\bar\vartheta_\RR$ do not rely on \cref{ass:rare}.  
\end{enumerate}

\end{theorem}

Generally, in both cases of strong ignorability and monotonicity, case-population sampling provides an easier environment for causal inference than case-control studies: $\Psi_{\CP,\RR}(0,0)$ does not depend on $p$ and $\Psi_{\CP,\AR}(p)$ is linear in $p$.  Also, the bounds under strong ignorability are all comparable with those under monotonicity: the upper bounds have the same form under strong ignorability as under monotonicity except that the monotonicity assumptions impose restrictions on the direction of the causal effect. 

\Cref{cor:aggregation1,cor:aggregation2} show that $\bar \vartheta_\RR$ suites better case-control or case-population studies than $\bar \vartheta_\AR$, especially when the case is potentially rare, despite the popularity of the latter in random sampling. Specifically, $r_\CC(x,p)$ and $r_\CP(x,p)$ should be taken into account for $\bar\vartheta_\AR$, but they are irrelevant for $\bar\vartheta_\RR$. This is an important difference because $r_\CC(X,0) = r_\CP(X,0) = 0$, which implies that the bounds on $\bar\vartheta_\AR$ cannot be tighter under strong ignorability than under monotonicity. In order to see the point more clearly, consider the case of case-control studies, i.e., \cref{design1}, and suppose that $\max_{p\in[0,\bar p]} \bigl\{ \Psi_{\CC,\AR}(p,1)p + \Psi_{\CC,\AR}(p,0)(1-p)  \bigr\} > 0$ so that the upper bound on $\bar\vartheta_\AR$ is positive both under strong ignorability and under monotonicity. In this case, the lower bound on $\bar\vartheta_\AR$ under strong ignorability can never be strictly positive because $\Psi_{\CC,\AR}(0,y)$ is trivially equal to zero. In other words, strong ignorability does provide a more informative environment than monotonicity but only in the sense that the former does not restrict the sign of $\bar\vartheta_\AR$. Once the sign of $\bar\vartheta_\AR$ is given, then there is nothing extra the strong ignorability assumptions offer relative to the monotonicity setup in understanding the average treatment effect.  The same is true for the case-population case, i.e., \cref{design2}. 

If we focus on $\bar\vartheta_\RR$, then the average of the log odds ratios, i.e., $\beta(y):= \Psi_{\CC,\RR}(0,y) = \Psi_{\CP,\RR}(0,y) = \Exp\{ \log \OR(X) \mid Y=y\}$ becomes the central object for estimation and inference. For instance, in \cref{design2}, all we need is $\beta(0)$, which can be interpreted as $\bar\vartheta_\RR$ itself or its sharp upper bound, depending on whether we assume strong ignorability or monotonicity, respectively. In \cref{design1}, if \cref{ass:MTR} is imposed, then $\Psi_{\CC,\RR}(p,y)$ can be shown to be decreasing in $p$, and therefore we have $\sC_{\CC,\RR}(p) \leq \beta(1)p + \beta(0) (1-p)$. Since the right-hand side is linear in $p$, we can easily conduct inference on $\bar\vartheta_\RR$ uniformly in $p\in [0,\bar p]$ by using $\beta(y)$, though this can be conservative. 

The log odds ratio $\log\OR(x)$ has been a popular measure of association in case-control studies, and $\beta(y)$ is an aggregation of it by using the identified distribution of $X$ given $Y=y$.  
%\citet{AAA} derive the semiparametric efficiency bound for $\Exp\{ \log \OR(X) \}$ and propose efficient estimators that allow for high-dimensional machine learning estimators in the first stage. When $X$ is low dimensional, there is a simple algorithm for efficient estimation of $\beta(y)$, which can be implemented by using standard software. In online appendix B2, we describe the algorithm.
\citet{AAA} establish the semiparametric efficiency bound for $\Exp\{ \log \OR(X) \}$ and suggest efficient estimators that accommodate high-dimensional machine learning estimators in the first stage. For low-dimensional $X$, a straightforward algorithm for efficient estimation of  $\beta(y)$ is available, and it can be easily implemented using standard software. The algorithm is described in online appendix B2.

\section{Causal inference under monotonicity}\label{sec:causal:inference}

In this section, we discuss how to carry out causal inference on the aggregated parameters $\bar\vartheta_\RR$ and $\bar\vartheta_\AR$ under the MTR and MTS assumptions: inference under strong ignorability can be done by the same principles. In our discussion below, $z(1-\alpha)$ will be the $1-\alpha$ quantile of the standard normal distribution. 

We first consider relative risk, for which we use $\exp(\bar\vartheta_\RR)$ as the parameter of interest: see our discussion right below \cref{eq:aggregated parameters}. Our basis for inference is \cref{cor:aggregation2}. Let $\beta(y) := \Psi_{\CP,\RR}(0,y) = \Exp\{ \log \OR(X) \mid Y=y\}$ for $y=0,1$. 

Inference is easier when we have a case-population sample: all we need is $\beta(0)$. Since we have $1\leq \exp(\bar\vartheta_\RR) \leq \exp\{ \beta(0) \}$ by \cref{cor:aggregation2}, a $1-\alpha$ confidence interval for $\exp(\bar\vartheta_\RR)$ can be constructecd by $\bigl[ 1,\ \exp\{ \hat\beta(0) + z(1-\alpha) \hat s(0)  \} \bigr]$, where $\hat\beta(0)$ is an asymptotically normal estimator of $\beta(0)$, and $\hat s(0)$ is its standard error. 

In the case of case-control sampling, i.e., \cref{design1}, we should base our inference on $\sC_{\CC,\RR}(\cdot)$. However, $\sC_{\CC,\RR}(p)$ is nonlinear in $p$, and hence it is difficult to obtain a confidence band uniformly in $p\in[0,\bar p]$. We propose two solutions.  
%One is just to use one-sided pointwise confidence bands using Efron's bias-corrected percentile intervals: for computational details, see algorithm 2 in online appendix D, where we focus on pointwise inference for $\bar\vartheta_\AR$, since computation for the two cases are similar. 
One is just to use one-sided pointwise confidence bands using bootstrap, akin to algorithm 2 in online appendix D, where we focus on pointwise inference for $\bar\vartheta_\AR$. 
The other is to take a conservative approach by using the fact that $\sC_{\CC,\RR}(p) \leq \tilde\beta(p) := \beta(1) p + \beta(0)(1-p)$. Specifically, in online appendix D, we show that 
\begin{equation}\label{def:ucb}
	\Pr \bigl[ 
		\forall p \in [0,1],\ 
		\exp \{ \tilde\beta(p) \} 
		\leq 
		\exp\{ p \hat\beta(1) +(1-p) \hat\beta(0) + u(1-\alpha) \}  
		\bigr] 
	\geq 
	1-\alpha,
\end{equation} 
where $u(1-\alpha) := z(1-\alpha/2) \max\{ \hat s(0),\ \hat s(1) \}$ with $\hat s(y)$ is the standard error of $\hat \beta(y)$, the asymptotically normal estimator of $\beta(y)$.

We now turn to inference on $\bar\vartheta_\AR$. The case-population sample provides an easier environment again: we can exploit the fact that $\Psi_{\CP,\AR}(p)$ is a simple linear function with the form of $\Psi_{\CP,\AR}(p) := p \xi_\CP$, where $\xi_\CP$ is implicitly defined here and does not depend on $p$. For more details, see online appendix D.

Inference on $\bar\vartheta_\AR$ with a case-control sample relies on the function $\sC_{\CC,\AR}(\cdot)$, and its nonlinearity in $p$ makes it difficult to construct a uniform confidence band. Since $\Psi_{\CC,\AR}(\cdot,y)$ is not monotonic, the conservative approach we discussed for $\bar\vartheta_\RR$ does not apply here. Therefore, we propose using one-side pointwise confidence intervals, for which we use Efron's bias-corrected percentile intervals. Computational details for implementation are given in online appendix D.

\section{Empirical Examples}\label{sec:application}

\subsection{Case-control sampling: entering a very selective university}\label{case-ctrl-example}

We consider quantifying the causal effect of attending private school on entering a very selective university by using the Pakistan data collected by \citet{Delavande:Zafar:19}. This is survey data from male students who were already enrolled in different types of universities in Pakistan, all located in Islamabad/Rawalpindi and Lahore. \citet{Delavande:Zafar:19} include two Western-style universities, one Islamic university, and four madrassas, but we focus on the two Western-style ones in our analysis: between the two universities, \citet{Delavande:Zafar:19} call the more expensive, selective, and reputable university "Very Selective University" (VSU) and the other simply "Selective University" (SU). Therefore, we restrict the population of interest to those who entered either VSU or SU, and we define the binary outcome to be whether a student entered VSU. The binary treatment we consider is whether a student attended private school before university.  Since the students in the sample were already enrolled in either VSU or SU at the time of the survey, we have a case-control sample, i.e. our \cref{design1}.

\begin{table}[htp]
	\centering
	\caption{University entrance and private school attendance \label{tb:univ-example}}
	\vspace{1ex}
	\hspace{-3.5cm}
	\begin{tabularx}{11cm}{XcX}
		&	
		\begin{tabular}{l|rrr}
			\hline\hline
			& \multicolumn{2}{c}{Private School} & Total \\
			University  & $T = 0$ &    $T = 1$ &  \\  \hline
			$Y = 0$ (SU) &    151 &   332 &  483   \\
			$Y = 1$ (VSU) &    51 &   155 &    206   \\  
			Total        &  202 &  487 &  689   \\
			\hline
		\end{tabular}
	\end{tabularx}
	%	\vspace{1ex}
\end{table}%

\Cref{tb:univ-example} shows the likelihood of entering VSU by private school attendance before university.
The empirical odds ratio is 1.38. 

In this example, the unconfoundedness assumption is unlikely to hold, because those who attended private school before university are likely to have more resourceful parents. This concern may not completely disappear even if we control for parental income and wealth because of the presence of unobserved parental abilities and resources that could affect their children's university choice. However, the MTR and MTS assumptions are still plausible: private school is probably no inferior input to university preparations (hence, MTR), and those who actually chose to attend private school probably care about their future college choice no less than those who did not (hence, MTS). Then, the odds ratio of 1.38 can be interpreted as a sharp upper bound on causal relative risk; therefore, the effect of attending private school seems, at best, modest.  

Now, we consider controlling for family background variables. Specifically, we include an indicator for at least one college-educated parent and parents' monthly income as covariates. \Cref{univ:est} reports estimation results for the aggregated log odds ratio within each of the case and the control: both $\beta(y)$ and $\exp\{\beta(y)\}$ convey the same information, but $\exp\{ \beta(y)\}$ is easier to interpret because it is comparable to the usual odds ratio in terms of its scale. The fact that $\widehat\beta(1)$ and $\widehat\beta(0)$ are notably different suggests that the amount of heterogeneity among individuals may be substantial. The confidence intervals are computed based on the MTR and MTS: hence, they are one-sided.

% latex table generated in R 3.6.2 by xtable 1.8-4 package
% 
\begin{table}[htbp]
	\caption{\label{univ:est} Estimation results of attending private school on entering VSU}\centering\medskip
	\begin{tabular}{lcc} \hline \hline
		& (1) & (2) \\
		& Case $(y=1)$  & Control $(y=0)$  \\  
		& VSU & SU \\
		\hline 
		$\beta(y)$ & 0.09 & 0.23 \\ 
		95\% confidence interval & [0,\,0.45] & [0,\,0.60] \\ 
		$\exp[\beta(y)]$ & 1.10 & 1.26 \\ 
		95\% confidence interval & [1,\,1.57] & [1,\,1.82] \\ 
		\hline \\
	\end{tabular}
	\parbox{5in}{\small{Note: Parental background is linearly controlled for when fitting retrospective binary logistic regression models.}} 
\end{table}

\begin{figure}[htbp]
	\begin{center}
		
		\caption{Causal inference on RR and AR: bounding the effects of attending private school on entering VSU}\label{fig_univ}
		
		\includegraphics[scale=0.35]{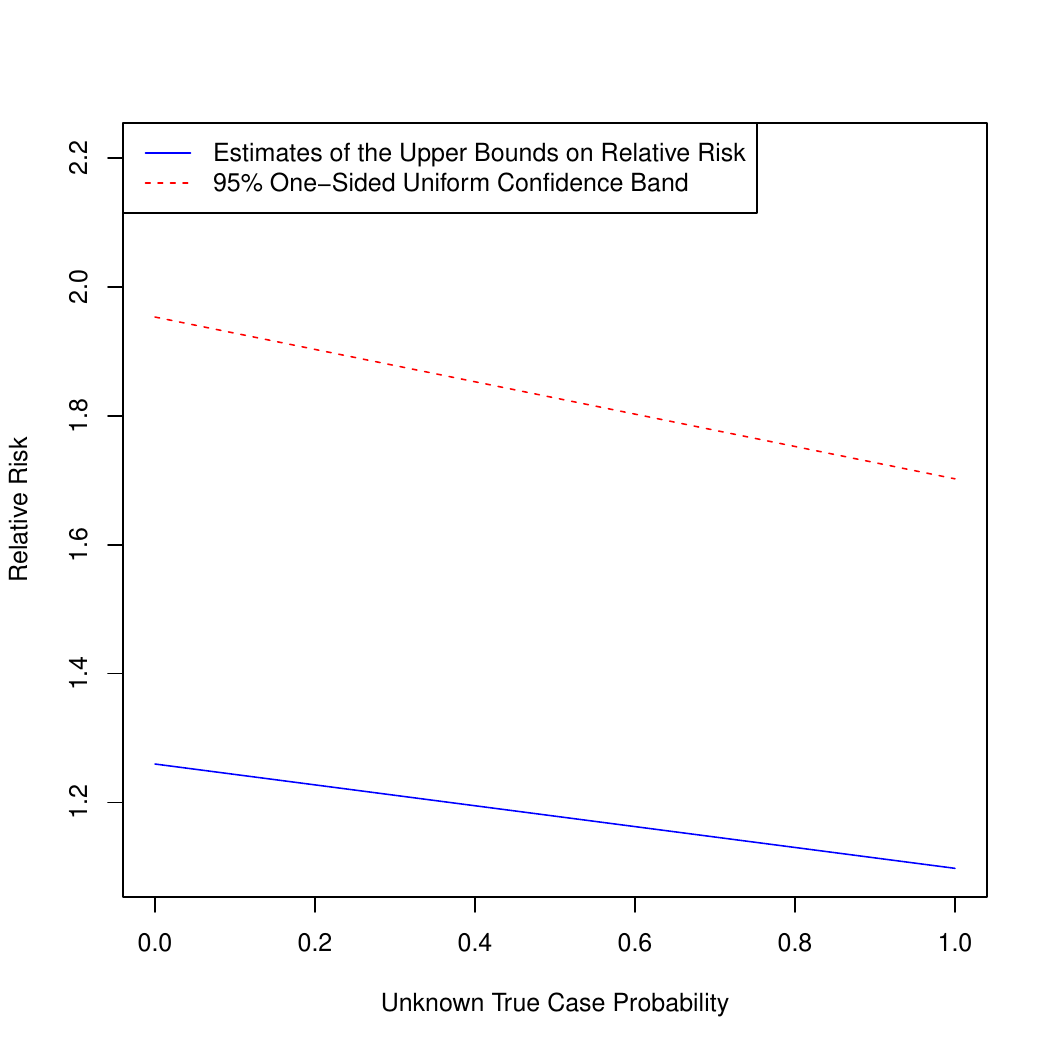}
		\includegraphics[scale=0.35]{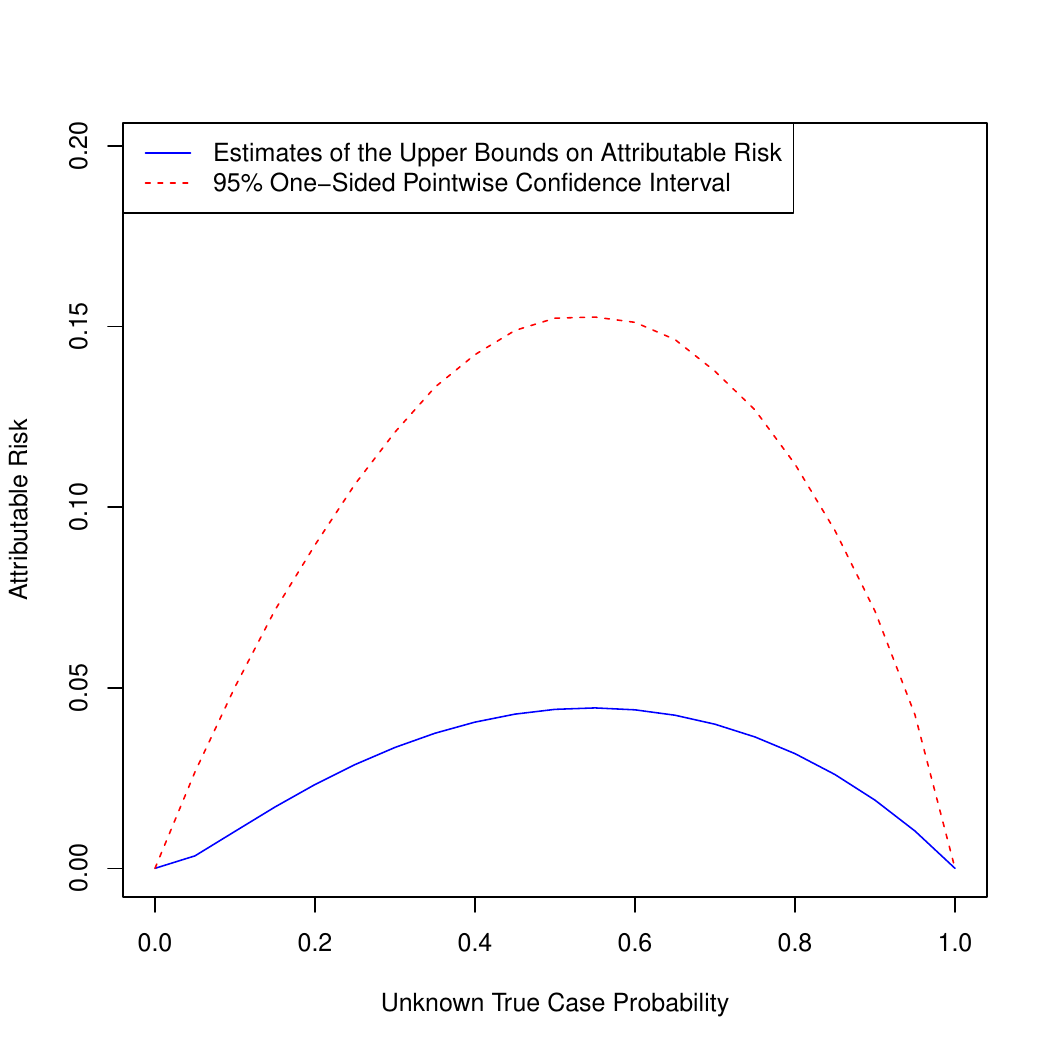}
		
	\end{center}
	\parbox{6in}{\small{
		Note:  
		The left panel shows the estimates and the 95\% one-sided uniform confidence band for the upper bounds on relative risk  and the right panel the estimates and the 95\% one-sided pointwise confidence intervals for the upper bounds on attributable risk, as functions of the unknown true case probability, i.e., $\Pr(Y^*=1)$.} }	
\end{figure}

We now consider the methods described in \cref{sec:causal:inference}, i.e., causal inference on the aggregated relative and attributable risk (RR and AR, respectively) in terms of the population distribution of the covariates. We rely on the MTR and MTS assumptions to interpret our results as upper bounds. For AR, we use the same covariates as in RR. The number of the bootstrap replications was 10,000. 

\Cref{fig_univ} summarizes the results: the left (right) panel shows RR (AR). The case probability $p_0$ of entering VSU in the population is not identified in this dataset. But we can trace out the upper bounds as the value of $p_0$ varies between $0$ and $1$. 

Consider the left panel of \cref{fig_univ}, i.e., RR, where we take the conservative approach and plot $\tilde\beta(p) = \beta(1) p + \beta(0)(1-p)$. If we take the point estimate at face value, attending private school increases the chance of entering VSU by a factor of at most 1.26. Even in terms of the confidence intervals, it seems highly unlikely that the impact is more than a factor of 2. The right panel of \cref{fig_univ} shows AR. The graph shows an inverted U-shape, because $r_\CC(x,p)\Gamma_{\CC,\AR}(x,p) = 0$ whenever $p$ is either $0$ or $1$. The maximum point estimate of the upper bound is 0.044, while the maximum value of the confidence intervals is 0.153. Therefore, it seems highly unlikely that attending private school increases the chance of entering VSU by more than 16 percent. 

None of our results require strong ignorability or the rare-disease assumption. Our conclusion of a relatively small positive effect of attending private school on entering VSU, if it exists at all, is reminiscent of existing results in labor economics that find access to private schools have only modest effects on children's performance \citep[see, e.g.,][]{Epple:17,MacLeod:19}.

\subsection{Case-population sampling: joining a criminal gang}\label{case-pop-example}

We revisit \citet{carvalho2016living}, who combine the 2000 Brazilian Census with a unique survey of drug-trafficking gangs in favelas (slums) of Rio de Janeiro; therefore, their dataset is an example of case-population sampling, i.e., our \cref{design2}. In their study, they use the method of \citet{lancaster1996case} to estimate a model of selection into the gang by using race, age, illiteracy, house ownership, and religiosity.  They note that the five characteristics are likely to be predetermined while years of schooling may be endogenous to entry, i.e., joining the gang may lead members to drop out of school. Indeed, 90 percent of gang members are not in school, whereas 46 percent of men aged 10--25 are not in school. They find that ``younger individuals, from lower socioeconomic background (black, illiterate, and from poorer families) and with no religious affiliation are more likely to join drug-trafficking gangs.''

\Cref{gangs:sum:stat} provides summary statistics of the sample. We regard \emph{currently not attending school} as the treatment variable of interest. Unconfoundedness is not plausible because of the endogeneity of schooling that we mentioned earlier. 
Furthermore, it is plausible that unmeasured factors such as family support could affect both treatment and outcome. 
However, not being in school may increase the chance of exposure to gang-related activities, and those who chose to be in school may be the ones who care for consequences no less than those who chose not to be; therefore, the MTR and MTS follow, respectively.

% matrix: data_stat file: gans_sum_table.tex  20 Aug 2020 09:32:50
\begin{table}[htbp]
	\caption{\label{gangs:sum:stat} Summary Statistics of the Case-Population Sample}\centering\medskip
	\begin{tabular}{lcc} \hline \hline
		& (1) & (2) \\
		& Case  & Population  \\  
		& Gang members & Men aged 10-25 \\
		\hline 
		Not in school &     0.901 &     0.458 \\  
		Black &     0.269 &     0.142 \\  
		Age &    16.722 &    17.526 \\  
		Illiterate &     0.094 &     0.041 \\  
		Owns house &     0.735 &     0.832 \\  
		No religion &     0.426 &     0.237 \\  
		\hline
		Sample size & 223 & 17175 \\
		\hline \\
	\end{tabular}
	\parbox{5in}{\small{Notes: Each entry shows the sample mean. Age is in years and all other variables are binary indicator variables. The population consists of men aged 10-25 living in Rio's Favelas.}} 
\end{table}

\Cref{gangs:est} presents estimation results for $\beta(y)$ and $\exp\{\beta(y)\}$, for which we control for the same covariates as \citet{carvalho2016living}. Unlike \cref{univ:est}, $y=0$ now corresponds to the entire population. Therefore, $\beta(0)$ itself is the log odds ratio aggregated over the population, which is the sharp upper bound on the aggregation of the log causal relative risk, i.e., $\bar\vartheta_\RR$.  The point estimate of $\beta(0)$ is 2.71, and that of $\exp\{\beta(0)\}$ is 15.01, which suggests that the chance of those who are not in school joining a gang may be (up to) 15 times as large as that of those who are.

% latex table generated in R 3.6.2 by xtable 1.8-4 package
% 
\begin{table}[htbp]
	\caption{\label{gangs:est} Estimation results of currently not attending school on relative risk of joining a gang}\centering\medskip
	\begin{tabular}{lcc} \hline \hline
		& (1) & (2) \\
		& Case  & Population  \\  
		& Gang members & Men aged 10-25 \\
		\hline 
		$\beta(y)$ & 2.90 & 2.71 \\ 
		95\% confidence interval & [0,\,3.36] & [0,\,3.19] \\ 
		$\exp[\beta(y)]$ & 18.10 & 15.01 \\ 
		95\% confidence interval & [1,\,28.90] & [1,\,24.39] \\ 
		\hline\\
	\end{tabular}
	\parbox{5in}{\small{Note: Race, age, illiteracy, house ownership, and religiosity are linearly controlled for when fitting retrospective binary logistic regression models.}} 
\end{table}

Our discussion above can be supplemented by checking the causal AR. At the three points of $p_0\in\{0.05,0.10, 0.15\}$ that \citet{carvalho2016living} considered, the point estimates and the end-points of the uniform confidence interval (in parentheses) for the upper bound on the causal AR are 0.33 (0.43), 0.66 (0.86), and 0.99 (1), respectively. The uniform confidence band is based on 1,000 bootstrap replications. Note that the confidence band is truncated at one, because AR cannot be larger than one.

Overall, our results are suggestive of potentially large impacts of keeping young men in school in order to discourage them to participate in criminal activities. Further research based on careful study designs would be necessary to reach a more definitive answer.

\subsection{Random sampling: physician's hours}\label{random-example}

\citet{FangGong17} construct estimates for physicians’ hours spent on Medicare beneficiaries and find that about 3 percent of physicians billed for more than 100 hours  per week. They refer to these physicians as \emph{flagged physicians}. 
\citet[p. 573]{FangGong17} state that ``flagged physicians are slightly more likely to be male, non-MD, more experienced, and provide fewer E/M services. Importantly, they work in substantially smaller group practices (if at all), and have fewer hospital affiliations.'' We use their study to illustrate the findings in this paper. Specifically, the  outcome variable is whether a physician billed for more than 100 hours per week in either 2012 or 2013, the treatment variable is a binary indicator showing whether or not the number of group practice numbers is less than 6, which is the median size in the data, and the covariates include an indicator for male, an indicator for doctor of medicine (MD), and experience in years (cubic polynomial).  

\begin{table}[htp]
	\centering
	\caption{Physician's potential overbilling and  size of the group practice\label{tb:FG-example}}
	\vspace{1ex}
	\hspace{-3.5cm}
	\begin{tabularx}{11cm}{XcX}
		&	
		\begin{tabular}{l|rrr}
			\hline\hline
			& \multicolumn{2}{c}{Small Group Practice} & Total \\
			Physician  & $T = 0$ &    $T = 1$ &  \\  \hline
			$Y = 0$ (Never flagged) &    38556 &   37348 &  75904   \\
			$Y = 1$ (Flagged in either year) &    583 &   1678 &   2261   \\  
			Total        &  39139 &  39026 &  78165   \\ 
			\hline
		\end{tabular}
	\end{tabularx}
	%	\vspace{1ex}
\end{table}%

The original dataset in \citet{FangGong17}  is updated in \citet{FangGong20} after \citet{Matsumoto} pointed out data and coding errors in the original work. In our analysis, we use the updated dataset.  \Cref{tb:FG-example} summarizes the sample, which consists of 78,165 physicians who billed at least 20 hours per week.

Treating this dataset as a random sample, we extract a case-control dataset:  
the case sample is composed of 2,261 flagged physicians;
the control sample of equal size is randomly drawn without replacement from the pool of physicians who were never flagged.
Analogously, a case-population dataset is obtained by combining the case sample with a population sample of equal size that is randomly drawn without replacement from all observations and its flagged status is coded missing.

It is highly unlikely that the group practice size is exogenous conditional on a small number of covariates; hence, we rely on the monotonicity assumptions (i.e., MTR and MTS) and focus on the upper bounds on the relative and attributable risks.

\begin{table}[htbp]
	\caption{\label{FG:est:RR}Odds Ratios}\centering\medskip
	\begin{tabular}{lccccc} \hline \hline
		& (1) & (2) & (3) & (4) & (5) \\
		& Random   &  \multicolumn{2}{c}{Case-Control} &  \multicolumn{2}{c}{Case-Population}  \\
	         & Sample  &  \multicolumn{2}{c}{Sample} &  \multicolumn{2}{c}{Sample}  \\    
		& All obs. & Case & Control & Case & Population \\
		\hline 
		$\exp[\beta(y)]$ &  2.68 & 2.92 & 2.65 & 2.66 & 2.57  \\ 
		95\% confidence interval & [1,\,2.93] & [1,\,3.30] & [1,\,2.96] &  [1,\,3.01] & [1,\,2.87] \\
		\hline \\
	\end{tabular}
	\parbox{5.8in}{\small{Note: An indicator for male, an indicator for doctor of medicine (MD), and experience in years (cubic polynomial) are controlled for when fitting logistic regression models.}} 
\end{table}

\Cref{FG:est:RR} reports the estimates of $\exp[\beta(y)]$ and their one-sided confidence intervals of $\exp[\beta(y)]$ for each sampling scheme. The lower bound is 1 because of the MTR assumption and 
averaging is done for a relevant population in each column.
Because the proportion of the flagged physicians is less than 0.03, we invoke the rare disease assumption and regard $\exp[\beta(y)]$ as an approximation to the upper bound on RR. 
Although there are some noticeable differences across different columns, the estimates are similar. This is consistent with the identification result that the price to pay is less for identification of RR when we move from random sampling to outcome-dependent sampling. The story is different if we focus on AR. 
\Cref{FG:est:AR} shows that the upper bounds on AR under outcome-dependent sampling are much larger than those under random sampling, especially when $\bar p = 0.1$.

\begin{table}[htbp]
	\caption{\label{FG:est:AR}Bounds on Attributable Risk}\centering\medskip
	\begin{tabular}{lccccc} \hline \hline
		& (1) & (2) & (3) & (4) & (5)  \\
		& Random   &  \multicolumn{2}{c}{Case-Control} &  \multicolumn{2}{c}{Case-Population}  \\
	         & Sample  &  \multicolumn{2}{c}{Sample} &  \multicolumn{2}{c}{Sample}  \\    
		&  & $\bar p=0.05$ & $\bar p=0.10$ & $\bar p=0.05$ & $\bar p=0.10$ \\
		\hline \\
		Bound estimate  &  0.024 & 0.044 & 0.083 & 0.044 & 0.088  \\ 
		95\% confidence interval & [0,\,0.027] & [0,\,0.051] & [0,\,0.095] &  [0,\,0.052] & [1,\,0.104] \\ 
		\hline\\
	\end{tabular}
	\parbox{5.8in}{\small{Note. The same covariates as in \cref{FG:est:RR} are controlled for. Each confidence interval is based on 1000 bootstrap replications.}} 
\end{table}

\bigskip
%\setstretch{1.2}
\setstretch{1}
\bibliography{casecontrol_SL} %\SJtodo{check Chen, X.}
\setstretch{\BASELINESTRETCH}

%\end{document}

%%%%%%%%%%%%%%%%%%%%%%%%%%%%%%%%%%%%%%%%%%%%%%%%%%%%%%%%%%%%%%%%%%%%%%%%%%%%%%%%%%%%%%%%%%%%%%%%%%%%%%%%%%%%%%%%%%%%%%%%%%%%%%

%%%%%%%%%%%%%%%%%%%%%%%%%%%%%%%%%%%%%%  ONLINE APPENDICES  %%%%%%%%%%%%%%%%%%%%%%%%%%%%%%%%%%%%%%%%%%%%%%%%%%%%%%%%%%%%%%%%%%%

%%%%%%%%%%%%%%%%%%%%%%%%%%%%%%%%%%%%%%%%%%%%%%%%%%%%%%%%%%%%%%%%%%%%%%%%%%%%%%%%%%%%%%%%%%%%%%%%%%%%%%%%%%%%%%%%%%%%%%%%%%%%%%

\clearpage
\appendix

\counterwithout{equation}{section}
\counterwithout{table}{section}
\counterwithout{figure}{section}

\counterwithout{lemma}{section}
\counterwithout{theorem}{section}
\counterwithout{assumption}{section}

\renewcommand{\thepage}{A-\arabic{page}}
\setcounter{page}{1}

\renewcommand{\theequation}{A.\arabic{equation}}
\setcounter{equation}{0}
\renewcommand{\thetable}{A.\arabic{table}}
\setcounter{table}{0}
\renewcommand{\thefigure}{A.\arabic{figure}}
\setcounter{figure}{0}

\renewcommand{\theassumption}{A.\arabic{assumption}}
\setcounter{assumption}{0}
\renewcommand{\thelemma}{A.\arabic{lemma}}
\setcounter{lemma}{0}
\renewcommand{\thetheorem}{A.\arabic{theorem}}
\setcounter{theorem}{0}

%%%%%%%%%%%%%%%%%%%%%%%%%%%%%%%%%%%%%%%%%%%%%%%%%%%%%%%%%%%%%%%%%%%%%%%%%%%%%%

\section*{\Large  Online Appendices to
	``Causal inference under outcome-based sampling with monotonicity assumptions''}

\begin{center}
	Sung Jae Jun \hspace{.2cm}
	\\
	Department of Economics, Pennsylvania State University\\
	and \\
	Sokbae Lee \\
	Department of Economics, Columbia University and IFS \\ 
	\vspace{1cm}
	October 19, 2023
	\vspace{2cm}
\end{center}

\begin{abstract}	
	\noindent
	In \cref{app:odds:ratio:no-log,sec:estimating betay,appx:MC}, we discuss aggregation of $\theta_\RR(\cdot)$ without taking the logarithm, efficient estimation of $\beta(y)$ for $y=0,1$, and the results of a small Monte Carlo experiment, respectively. \Cref{appx:algo} gives details on inference issues omitted in \cref{sec:causal:inference}, and \cref{appx:proofs} presents proofs.
\end{abstract}

\vspace{1cm}
%\todo[inline]{Created a cover for the appendix, because it will be a separate document. Lemma numbering was messed up in the appendix: we numbered everything in the appendix in the format of ``Lemma A.Number'' but LaTex automatically reset the counter in each section in the appendix. As a result, we had a few lemmas with the same numbering. Corrected all numberings (lemmas, theorems, etc) in the appendix.}

\clearpage
\section{Averaging without taking the logarithm}\label{app:odds:ratio:no-log}

In the main text we followed the convention of using the logarithm of the odds ratio, and we focused on aggregated versions of the logarithm of the odds ratio, i.e., $\beta(y) = \Exp\bigl[ \log\{\OR(X)\} \mid Y=y  \bigr]$ for $y=0,1$. As a result, the aggregated version of the central causal parameter in the main text was the aggregated logarithm of relative risk, i.e., $\bar \vartheta_\RR := \int_{\mathcal{X}} \log\bigl\{ \theta_\RR(x) \bigr\} f_{X^*}(x)\ dx$. 

Alternatively, one may want to proceed without taking the logarithm, which does not change the substance of our results, though the aggregate parameter of interest is now $\overline\zeta_\RR := \Exp \bigl[ \theta_\RR( X^* )  \bigr]$, for which we rely on  
\begin{equation*}
\zeta_\RR (y) := \int_\mathcal{X}  \theta_\RR(x)  dF_{X|Y}(x|y),\ \;
\kappa_\RR (y) := \int_{\mathcal{X}}  \OR(x) dF_{X|Y}(x|y)
\end{equation*}
for $y=0,1$.  Again, if the MTR and MTS conditions are satisfied, then we have
\begin{equation}\label{eq:bounds:no-log}
1\leq \zeta_\RR(y) \leq \kappa_\RR(y)
\end{equation}
under both \cref{design1,design2}, where the inequalities are sharp. We can construct efficient estimators of $\kappa_\RR (y)$ and carry out causal inference on $\overline\zeta_\RR$ by using the same idea as we described in \cref{sec:estimation}. We do not repeat all the details for brevity. 

In general, we have $\Exp\{ \log\OR(X^*) \}\leq \log\Exp\{\OR(X^*) \}$ by Jensen's inequality. Therefore, an average of the log odds ratio is less likely to be affected unduly by outliers than that of the odds ratio itself. This seems to be another merit in using the logarithm of the odds ratio in addition to the usual advantage that it corresponds to the coefficients of the treatment variable when a parametric logistic model is used.

\section{Efficient Estimation of \texorpdfstring{$\beta(y)$}{beta(y)}} \label{sec:estimating betay}
In this section, we discuss efficient estimation of $\beta(y) := \Exp\bigl[ \log\{\OR(X)\} \mid Y=y  \bigr]$ for $y=0,1$. For this purpose, we point out that the sample consists of independent and identically distributed copies of $(Y,T,X)$: the sampling designs affect the distribution of $(Y,T,X)$ and its relationship with that of $(Y^*,T^*,X^*)$.  Therefore, all regularity conditions and results in this section will be presented in terms of the observed variables $(Y,T,X)$, and hence the regularity conditions are testable in principle.

\subsection{The efficient influence function}\label{sec:eff:influence}

We first derive the semiparametric efficiency bound. We start with the following assumptions for regularity.

\begin{assumption}[Bounded Probabilities]\label{ass:DML}
	There is a  constant $\varepsilon > 0$  such that for each $y=0,1$,  $\varepsilon \leq \Pr(T=1\mid X,Y=y) \leq 1 - \varepsilon$ and $\varepsilon \leq \Pr(Y=1\mid X) \leq 1 - \varepsilon$ almost surely.
\end{assumption}

\begin{assumption}[Regular Distribution]\label{ass:regular-distn}
	The distribution function $F_{X|Y}$ has a probability density $f_{X|Y}$ that satisfies $0 < f_{X|Y}(x\mid y) < \infty$ for all $x\in \mathcal{X}$ and $y=0,1$. 
\end{assumption}

%\Cref{ass:DML,ass:regular-distn} are, in principle, testable since they are about the random variables observed in the sample. Therefore, \cref{ass:DML} is not at odds with \cref{ass:rare}. For example, under both \cref{design1,design2}, \cref{ass:DML} requires that $h_0 f_{X^*|Y^*}(x\mid 1)/f_{X^*}(x)$ be bounded away from zero. 

\Cref{ass:DML} is slightly stronger than what we need to derive the efficient influence function, but it is useful to ensure that all the population quantities given below are well-defined without spelling out all the conditions. \Cref{ass:regular-distn} focuses on the case where $X$ is continuous, but this is only for the sake of notational simplicity.

%if $X$ is discrete or mixed, then $f_{X|Y}$ should be understood as a general Radon-Nikodym density with respect to some dominating measure.  

Under Bernoulli sampling, the likelihood of a single observation $(Y,T,X)$ can be expressed as a mixture of two binary likelihoods, from which we first have the following lemma.

\begin{lemma}\label{lem:tangent}
	Consider the Bernoulli sampling scheme of \cref{design1} or \cref{design2}. The tangent space can be represented by the set of functions of the following form:
	\begin{multline*}
	s(Y,T,X) = 
	(1-Y)\Bigl[ a_0(X) + \bigl\{ T- \Pr(T=1\mid X,Y=0)  \bigr\} b_0(X)  \Bigr] \\
	+
	Y \Bigl[ a_1(X) + \bigl\{ T- \Pr(T=1\mid X,Y=1) \bigr\} b_1(X)  \Bigr],
	\end{multline*}	
	where the functions $a_y$ and $b_y$ are such that $\Exp\{ a_y(X) \mid Y=y \} = 0$ and $\Exp\{ s^2(Y,T,X) \}<\infty$  for each $y=0,1$.
\end{lemma}

By direct calculation, as in \citet{hahn1998role}, we can show that $\beta(y)$ is pathwise differentiable along regular parametric submodels in the sense of \citet{newey1990semiparametric,newey1994asymptotic}. Further, it turns out that the pathwise derivative is an element of the tangent space presented in \cref{lem:tangent}, from which we can obtain the semiparametric efficiency bound for $\beta(y)$.  The following theorem presents this result. Let 
\begin{equation}\label{def:w_x}
w(X) := \frac{f_{X|Y}(X\mid 0)}{f_{X|Y}(X\mid 1)} = \frac{h_0}{1-h_0} \frac{\Pr(Y=0\mid X)}{\Pr(Y=1\mid X)},
\end{equation}
where the second equality is by the Bayes rule. Further, for $y=0,1$, define
\begin{equation*}
\Delta_y (Y,T,X) 
:= 
\frac{ Y^y (1-Y)^{1-y} \{T-\Pr(T=1\mid X,Y=y)\}}{\Pr(T=1\mid X,Y=y)\{1-\Pr(T=1\mid X,Y=y) \}}.
\end{equation*}

\begin{theorem}\label{thm:eff}
	Suppose that \cref{ass:support,ass:DML,ass:regular-distn} hold and that we have a sample by Bernoulli sampling. Then, for $y=0,1$, $\beta(y)$ is pathwise differentiable and its pathwise derivative is given by
	\begin{multline*}
	F_y(Y,T,X) 
	=
	\frac{Y^y(1-Y)^{1-y}}{h_0^y (1-h_0)^{1-y}}
	\Bigl\{
	\log \OR(X) - \beta(y)	\Bigr\} \\
	-
	\frac{\Delta_0 (Y,T,X)}{(1-h_0)w(X)^y} 
	+
	\frac{w(X)^{1-y}  \Delta_1 (Y,T,X)}{h_0}.  
	\end{multline*}
	Further, $F_y$ is an element of the tangent space, and therefore, the semiparametric efficiency bound for $\beta(y)$ is given by $\Exp\bigl\{ F_y^2(Y,T,X) \bigr\}$. 
\end{theorem}

\Cref{thm:eff} implies that the asymptotic variance of a $\sqrt{n}$--consistent and asymptotically linear estimator of $\beta(y)$ should be
at least $\Exp\{ F_y^2(Y,T,X) \}$ by Theorem 2.1 of \citet{newey1994asymptotic}. The first term that appears in $F_y(Y,T,X)$ has mean zero, because
\begin{equation}\label{eq:gmm} 
\Exp\Biggl[ \frac{Y^y(1-Y)^{1-y}}{h_0^y (1-h_0)^{1-y}}  \Bigl\{ \log \OR(X) - \beta(y)	\Bigr\} \Biggr] 
=
\Exp\bigl\{ \log \OR(X) - \beta(y) \  \big| \ Y=y  \bigr\} 
= 0.
\end{equation}
The other terms in $F_y(Y,T,X)$ are for adjustment to address the effect of first step nonparametric estimation of $\log \OR(X)$ via $\Pr(T=1\mid X=x,Y=y)$.  

We remark that the efficiency bound does not change even if $h_0$ is unknown, as long as the distribution of $(T,X)$ given $Y=y$ does not depend on $h_0$. This point can be seen as follows. Suppose that $h_0$ is unknown and that the distribution of $(T,X)$ given $Y=y$ does not depend on $h_0$. Then, the log-likelihood of a regular parametric submodel as we perturb the conditional distributions of $T,X$ given $Y=y$ can be written as
\[
\ell(h,\gamma; Y,T,X)  
= 
Y\{ \log h + \ell_1(\gamma; T,X) \} + (1-Y)\{ \log(1-h) + \ell_0(\gamma; T,X)  \},   
\]
where $\ell_y(\gamma;T,X) = \log \Pr(T,X|Y=y; \gamma)$ is the log-likelihood of the conditional distribution of $T,X$ given $Y=y$ along a parametric perturbation represented by $\gamma$ (with the true vale at $\gamma_0$). Here, $\ell_y(\gamma; T,X)$ does not depend on $h$ by assumption. Therefore, the information matrix is block diagonal, and the score along the $\gamma$--dimension in this setup is the same as the score under the assumption that $h_0$ is known.  Since the new tangent space is obtained by a linear span of the scores along the $h$ and $\gamma$ dimensions, it is always a larger space than the tangent space we obtained under the assumption that $h_0$ is known. 

Now, note that the parameter $\beta(y)$ depends only on $\Pr(T,X\mid Y=1), \Pr(T,X\mid Y=0)$, and $f(X\mid Y=y)$, none of which depend on $h_0$ by assumption. Therefore, only the perturbation along the $\gamma$--dimension will affect $\beta(y)$, which means that the perturbed parameter $\beta(y; h, \gamma)$ along $(h,\gamma)$ will be just the same as $\beta(y; \gamma)$. Therefore, the pathwise derivative of $\beta(y)$ stays the same whether $h_0$ is known or not. 

Since we already showed that the pathwise derivative of $\beta(y)$ is contained in the tangent space we obtained under the assumption that $h_0$ is known, it must be included in the new tangent space as well. Therefore, the pathwise derivative itself is again the efficient influence function, which is the same whether $h_0$ is known or not. This point can be seen from a slightly different angle based on the moment condition and block diagonality of the information. See \cref{sec:estimation} for more details.  

\subsection{An efficient estimation algorithm}\label{sec:estimation}

Efficient estimators of $\beta(y)$ for $y=0,1$ can be constructed in multiple ways. The most straightforward approach is using the moment condition
\begin{equation}\label{eq:m_y}
	\Exp\left\{ \frac{Y^y(1-Y)^{1-y} \{ \log \OR(X) - \beta(y) \} }{h_0^y (1-h_0)^{1-y}}  \right\} = 0, 
\end{equation}
where we plug in a nonparametric estimator of $\OR(x)$. A resulting estimator will be $\sqrt{n}$-consistent, asymptotically linear, and semiparametrically efficient under suitable conditions.  Below we propose a simple algorithm that implements this idea, for which we use sieve logistic estimators for the first step nonparametric estimation. 

Suppose that we have an i.i.d.\ sample, $\bigl\{ (Y_i,T_i, X_i): i=1,\dots,n \bigr\}$: this is not an i.i.d.\ sample of $(Y_i^*, T_i^*, X_i^*)$. Throughout the discussion, we assume that $h_0$ is known; otherwise, we can use $\hat h := \sum_{i=1}^n Y_i /n$ instead, which does not affect the asymptotic variance (or equivalently the efficiency bound), provided that the distribution of $(T,X)$ given $Y=y$ does not depend on $h_0$. Indeed, if $\OR(x)$ does not depend on $h_0$, then we can index the moment condition in \cref{eq:m_y} along parametric submodels as $\Exp\{ m_y(h_0, \gamma_0, \beta(y); Y,X) \} = 0$, where 
\[
m_y(h, \gamma, \beta; Y,X):= \frac{Y^y(1-Y)^{1-y} \{ \log \OR(X; \gamma) - \beta \} }{h^y (1-h)^{1-y}}. 	
\]
Then, some simple algebra shows that $\Exp[ \partial m_y\{ h,\gamma_0, \beta(1); Y,X \} /\partial h |_{h=h_0} ] = 0$. Therefore, it is only the estimation of $\gamma_0$, not the estimation of $h_0$, that matters for the variance of the estimator of $\beta(y)$. But we already argued at the end of \cref{sec:eff:influence} that the information along $h$ and $\gamma$ is block diagonal. 

In order to estimate $\OR(x)$ nonparametrically in the first step, we consider infinite dimensional (retrospective) logistic regression: i.e., for $y=0,1$,
\begin{equation}\label{eq:retro}
\Pr(T=1 \mid X=x,Y=y)
=\frac{\exp\left\{ \sum_{j=1}^\infty \phi_j(x) \mu_{j,y}\right\}}{1+ \exp\left\{ \sum_{j=1}^\infty \phi_j(x) \mu_{j,y}\right\}},
\end{equation}
where $\{ \phi_j: j=1,2,\ldots \}$ is a series of basis functions and $\{ \mu_{j,y}: j=1,2,\ldots \}$ is a series of unknown coefficients for each $y=0,1$.  It then follows that $\log\OR(x)
=
\sum_{j=1}^\infty \phi_j(x)(\mu_{j,1} - \mu_{j,0})$,
%\begin{align}\label{retrospective-expansion}
%\log\OR(x)
%=
%\sum_{j=1}^\infty \phi_j(x)(\mu_{j,1} - \mu_{j,0}),
%\end{align}
from which we obtain
\begin{equation}
\beta (y)
\approx 
\sum_{j=1}^{J_n} 
\int_{\mathcal{X}} \phi_j(x) dF_{X|Y}(x|y)
\left( \mu_{j,1} - \mu_{j,0} \right), \label{sieve-approximation}
\end{equation}
provided that $J_n$ diverges as $n$ increases. \Cref{sieve-approximation} suggests the following two-step sieve estimation strategy:
\begin{enumerate}
	\item In the first step, for each $y=0,1$, estimate $\{ \mu_{j,y}: y=0,1 \; j=1,\ldots, J_n \}$ by logistic regression of $T_i$ on $\{ \phi_j(X_i): j=1,\ldots, J_n \}$ with the  $Y_i = y$ sample. 
	
	\item In the second step, construct a sample analog of \cref{sieve-approximation}: i.e.,
	\begin{equation}\label{retro-logit}
	\widehat{\beta}(y) := \sum_{j=1}^{J_n} 
	\int_{\mathcal{X}} \phi_j(x) d \widehat{F}_{X|Y}(x|y)
	\left( \widehat{\mu}_{j,1} - \widehat{\mu}_{j,0} \right),
	\end{equation} 
	where $\widehat{\mu}_{j,y}$'s are sieve logit estimates from the first step, and
	\begin{equation*}
	\int_{\mathcal{X}} \phi_j(x) d \widehat{F}_{X|Y}(x|y)
	= 
	\frac{\sum_{i=1}^n Y_i^y (1-Y_i)^y \phi_j(X_i)}{\sum_{i=1}^n Y_i^y (1-Y_i)^y}.
	\end{equation*}
\end{enumerate}
Since we use retrospective regression as shown in \cref{eq:retro}, we call
the estimator defined in \eqref{retro-logit} the \emph{retrospective sieve logistic estimator} of $\beta(y)$ for $y=0,1$. It can be computed using standard software for logistic regression,  as described in \cref{sieve-est-retro}.

\begin{algorithm}[h]
	\KwInput{$\{ (Y_i, T_i, X_i): i=1,\ldots,n \}$, tuning parameter $J_n$ and basis functions $\{ \phi_j(\cdot): j=1,\ldots, J_n \}$}
	\KwOutput{estimate of $\beta (1)$ and its standard error}
	
	Construct $\{ \phi_1 (X_i), \ldots, \phi_{J_n} (X_i): i=1,\ldots, n \}$, where an intercept term is excluded in $\phi_j$'s\;
	
	For each $j = 1,\ldots, J_n$, compute the empirical mean of  $\phi_j (X_i)$ using only the case sample ($Y_i = 1$) and construct the demeaned version, say $\varphi_j(X_i)$, of $\phi_j (X_i)$\;
	
	Run a logistic regression of $T_i$ on the following regressors:
	an intercept term, $Y_i$, $\varphi_j(X_i), j=1,\ldots,J_n,$ and interactions between
	$Y_i$ and $\varphi_j(X_i), j=1,\ldots,J_n,$ using standard software;
	
	Read off the estimated coefficient for $Y_i$ and its standard error
	
	\caption{Retrospective Sieve Logistic Estimator of $\beta (1)$}\label{sieve-est-retro}
\end{algorithm}

The procedure described in \cref{sieve-est-retro} achieves the first step by running a combined logistic regression of $T_i$ on $Y_i$, the sieve basis terms and the interactions between $Y_i$ and  the sieve basis terms. This is first-order equivalent since $Y_i$ is binary and full interaction terms are included. For the second step, instead of evaluating the right-hand side of \cref{retro-logit} after logistic regression, $\phi_j (X_i)$'s are demeaned first using only the case sample so that the resulting coefficient for $Y_i$ is first-order equivalent to the estimator defined in \cref{retro-logit}. The advantage of the formulation in \cref{sieve-est-retro} is that the standard error of  $\widehat{\beta}(1)$ can be read off directly from standard software without any further programming. 
It is straightforward to modify \cref{sieve-est-retro} for estimating $\beta (0)$. One has to compute the empirical mean of  $\phi_j (X_i)$ using only the control sample ($Y_i = 0$) for the demeaning step. 

It is not difficult to work out formal asymptotic properties of our proposed sieve estimator in view of the well-established literature on two-step sieve estimation \citep[see, e.g., ][among many others]{Ai2003,Ai2012,Ackerberg2014}. Since this is now well understood in the literature, we omit details for brevity of the paper. \Cref{appx:MC} reports the results of a small Monte Carlo experiment that illustrates the finite-sample performance of the proposed estimators of $\beta(1)$ and $\beta(0)$.

\section{Monte Carlo Experiments}\label{appx:MC}

In this section, we report the results of a small Monte Carlo experiment. A case-control sample is generated from 
\begin{equation*}
X \mid Y=y 
\sim 
\mathbb{N} (\mu^{(y)}, \Sigma^{(y)} )
\quad\text{and}\quad
\Pr(T = 1 \mid X=x, Y=y) 
= 
G( \alpha_0^{(y)}  + X^\tr \alpha_1^{(y)}),
\end{equation*}
where $G(u) = \exp(u)/\{1+\exp(u)\}$, $\alpha_0^{(y)}$, $\alpha_1^{(y)}$, $\mu^{(y)}$ and $\Sigma^{(y)}$ are parameters that may depend on $y=0,1$. In simulations we focus on estimating $\beta(y)$ that can now be expressed as $\beta(y) = ( \alpha_0^{(1)} - \alpha_0^{(0)} ) 
+ \Exp(X  | Y = y)^\tr (\alpha_1^{(1)} - \alpha_1^{(0)})$. 

With the dimension of $X$ equal to $d_x = 5$, the parameter values are specified as follows: 
$\mu^{(1)} = (1,\ldots,1)^\tr$,
$\mu^{(0)} = (0,\ldots,0)^\tr$, and
$\Sigma^{(y)} = \Sigma$ for $y=0,1$, where 
the $(j,k)$ element of $\Sigma$ is $\Sigma_{j,k} = \rho^{|j-k|}$ and $\rho = 0.5$;
$\alpha_0^{(1)} = 0.5$,
$\alpha_1^{(1)} = (1, 1, 0, 0, 0)^\tr$,
$\alpha_0^{(0)} = 0$,
$\alpha_1^{(0)} = (0, 0, 1, 1, 0)^\tr$. 
In this design we have $\beta(1) = \beta(0) = 0.5$.

%We conduct several sets of Monte Carlo experiments. 
In each Monte Carlo replication, we simulate 1,000 observations separately for both $Y=0$ and $Y=1$ samples (that is, the total sample size is 2,000 and $\hat{h} = 0.5$). 
There are 1,000 Monte Carlo replications.

\begin{table}[htbp]
	\centering
	\caption{Results of Monte Carlo Experiments \label{tb:MC}}
	\vspace{1ex}
	\begin{tabular}{lcccc}
		\hline
		& \multicolumn{2}{c}{$\beta(1)$} & \multicolumn{2}{c}{$\beta(0)$} \\ 
		& parametric & sieve   & parametric & sieve  \\ 
		\hline
		Mean Bias & 0.011 & 0.070 & 0.005 & 0.046 \\ 
		Median Bias & 0.012 & 0.086 & -0.001 & 0.042 \\ 
		RMSE & 0.057 & 0.167 & 0.033 & 0.067 \\ 
		Mean AD & 0.191 & 0.330 & 0.145 & 0.206 \\ 
		Median AD & 0.160 & 0.283 & 0.119 & 0.173 \\ 
		Cov. Prob. & 0.944 & 0.962 & 0.952 & 0.962 \\  
		\hline
	\end{tabular}
	\vspace{1ex}
	\par
	\parbox{5in}{Note: RMSE stands for the root mean squared error and AD refers to absolute deviation.
		Cov. Prob. is the coverage probability of the one-sided 95\% confidence interval. 
		The results are based on 1,000 Monte Carlo repetitions.}	
\end{table}%

We consider two estimators: (i) a retrospective parametric logistic estimator
that uses $X$ as covariates and (ii)  a retrospective sieve logistic estimator
that uses the linear, quadratic and interaction terms of $X$ as covariates (that is, $2d_x + d_x(d_x-1)/2 = 20$ covariates all together). 
\Cref{tb:MC} summarizes the results of the Monte Carlo experiments.
Not surprisingly, the parametric estimator performs better for both $\beta(1)$ and $\beta(0)$. It shows almost no bias
and small root mean squared errors and absolute deviations.
Its coverage probability is close to the nominal 95\%. 
The sieve estimator exhibits some positive biases but its performance is overall satisfactory.

\section{Inferential Details}\label{appx:algo}

In this section, we provide details that are omitted in \cref{sec:causal:inference}. We first show \cref{def:ucb}. Let $\bar\beta(y) :=\hat\beta(y) - \beta(y)$ for $y=0,1$, where $\hat\beta(y)$ is an asymptotically normal estimator of $\beta(y)$ with the standard error equal to $\hat s(y)$. Note that 
\[
    \mathbb{B} 
    := \inf_{p\in[0,1]} \bigl\{ p\bar\beta(1) + (1-p)\bar\beta(0)\bigr\}
    = 
    \min\{\bar\beta(1),\bar\beta(0) \}. 
\] 
Therefore, 
\[
    \Pr\{ \mathbb{B} \leq -u(1-\alpha) \} 
    \leq 
    \sum_{y=0}^1 \Pr\{ \bar \beta(y) \leq - u(1-\alpha)\}
    \leq 
    \alpha/2 + \alpha/2.
\] 
This procedure can be easily modified when \cref{ass:rare} with $\bar p<1$ is adopted. 

We remark that the asymptotic coverage rate in \cref{def:ucb} is conservative. Achieving the asymptotically exact coverage rate requires that we use the limiting distribution of $\min\{ \bar\beta(1), \bar\beta(0)  \}$, which is not normal and is not readily available from the estimation algorithm given in \Cref{sec:estimation}. Also, if one wants to directly use the nonlinear function $\sC_{\CC,\RR}(p)$, then the bootstrap-based algorithm described in \cref{ciar-cc} can be used: \cref{ciar-cc} focuses on $\sC_{\CC,\AR}$, but its modification for $\sC_{\CC,\RR}$ is straightforward.

\begin{algorithm}[htbp]
	\KwInput{$\{ (Y_i, T_i, X_i): i=1,\ldots,n \}$, 
		%prospective estimates of $\Pr(Y=1\mid X_i)$,
		%retrospective estimates of $\Pi(t\mid y,X_i)$ for $t =0, 1,\ y = 0, 1$,
		the number ($B$) of bootstrap replications, the coverage probability ($1-\alpha$) of the confidence interval, the upper bound ($\bar p$) on the unknown true case probability}
	
	\KwOutput{point estimates $\widehat{\sC}_{\CC,\AR}(p)$ of the upper bounds on causal attributable risk and the upper end points of the one-sided pointwise bootstrap confidence intervals
		$q_{(1-\alpha)}^*(p)$ for $p \in [0, \bar p]$}
	
	\vspace*{2ex}
	
	Construct a grid $\mathcal{P} := \{p_0,p_1,\ldots,p_J\}$ of $[0, \bar p]$, 
	where $0 = p_0 < p_1 < \dots < p_J = \bar p$;

	For each $p \in \mathcal{P}$, evaluate sample analogs $\widehat{\sC}_{\CC,\AR}(p)$ of $\sC_{\CC,\AR}(p) := p \Psi_{\CC,\AR}(p,1) + (1-p) \Psi_{\CC,\AR}(p,0)$; in this step, we need to compute retrospective estimates of $\Pi(t\mid y, X_i)$ for $t=0,1,\ y=0,1$ as well as prospective estimates of $\Pr(Y=1\mid X_i)$, because $\Psi_{\CC,\AR}(p,y)$ depends on $r_\CC(x,p)$ as its definition in \cref{def rcc} shows; 
	% (this step gives the point estimates of the upper bound on causal relative risk);
	
	For each bootstrap replication $b=1,\ldots,B$, generate a bootstrap sample 
	$\{ (Y_i^{*,b}, T_i^{*,b}, X_i^{*,b}): i=1,\ldots,n \}$, and obtain a bootstrap estimate
	$\widehat{\sC}^{*,b}_{\CC,\AR}(p)$ for each $p \in \mathcal{P}$;
	
	For each $p \in \mathcal{P}$, compute 
	$$\mu^*(p) := \frac{1}{B} \sum_{b=1}^B 	 \one \left\{ \widehat{\sC}^{*,b}_{\CC,\AR}(p) 
	\leq \widehat{\sC}_{\CC,\AR}(p) \right\},$$ where $\one\{ \cdot \}$ is the usual indicator function;
	
	For each $p \in \mathcal{P}$, obtain $\nu^*(p) := \Phi \left[ \Phi^{-1} (1-\alpha) + 2 \Phi^{-1} \{\mu^*(p) \} \right]$, where $\Phi(\cdot)$ is the cumulative distribution function of the standard normal random variable;
	
	For each $p \in \mathcal{P}$, compute the $\nu^*(p)$ empirical quantile of $\widehat{\sC}^{*,b}_{\CC,\AR}(p)$, say $q_{(1-\alpha)}^*(p)$, as the $(1-\alpha)$ one-sided pointwise bootstrap confidence interval
	
	\caption{Causal Inference on $\bar\vartheta_\AR$ Using Case-Control Samples}\label{ciar-cc}
\end{algorithm}

We now turn to $\bar\vartheta_\AR$. In the case-population case, we use the fact that $\Psi_{\CP,\AR}(p)$ has a simple linear form such as $\Psi_{\CP,\AR}(p) := p \xi_\CP$, where
\begin{align*}
 \xi_\CP &:= \Exp \biggl\{ \frac{(1-h_0)}{h_0}\frac{\Pr(Y=1\mid X=x)}{\Pr(Y=0\mid X=x)} \Gamma_{\CP,\AR}(X,0) \ \Big| \ Y=0 \biggr\}, \\
\Gamma_{\CP,\AR}(x,0) &=
	\sum_{j=0}^1
	\frac{(-1)^{j+1}\Pi(j\mid 1,x)}{\Pi(j\mid 0,x)}.	
\end{align*}
Therefore, a one-side confidence band that is uniform in $p$ can be constructed as
\[
    p\in[0,1] \mapsto \bigl[ 0,\ p\{\hat \xi_\CP + u_{\AR,\CP}(1-\alpha)\} \bigr],
\] 
where $\hat\xi_\CP$ is an estimator of $\xi_\CP$, and $-u_{\AR,\CP}(1-\alpha)$ is, for instance, the bootstrap $\alpha$-quantile of $\hat\xi_\CP - \xi_\CP$. We note here that the asymptotic validity of the nonparametric bootstrap in two-step semiparametric models is a well-studied topic \citep[e.g.,][]{chen2003estimation}.  However, in practice, the na\"{i}ve bootstrap may suffer from a finite sample bias, because ratios of probabilities need to be estimated in the first step. To mitigate this issue, we recommend using \citet{efron1982book}'s bias-corrected one-sided percentile interval to obtain $u_{\AR,\CP}(1-\alpha)$: see the bootstrap procedure we describe in \cref{ciar-cc}.  Finally, it is straightforward to adopt \cref{ass:rare} by restricting attention to $p\in [0,\bar p]$.

Computational details for the case of case-control sampling are given in \cref{ciar-cc}.

\section{Proofs}\label{appx:proofs}

It is worth noting that our big picture arguments consist of two parts. In Part 1, we suppose that the distribution of $(Y^*, T^*, X^*)$ were known, and we obtain the sharp identified intervals on the parameters of interest in terms of the distribution of $(Y^*,T^*,X^*)$. In Part 2, we show that the bounds obtained in Part 1 can be equivalently expressed in terms of the distribution of $(Y,T,X)$ and $p_0 = \Pr(Y^*)$.  Therefore, finding the final sharp identified bounds in terms of the distribution of $(Y,T,X)$ becomes just a matter of dealing with the unknown value of $p_0$, which can be taken care of by a simple continuity argument.

In this section, we present all proofs. 

\subsection{Auxiliary Results}\label{appx:lemmas}

 We first summarize all the restrictions imposed by \cref{ass:MTR,ass:MTS}. The conditional probability mass function of $\bigl( Y^*(0), Y^*(1), T^* \bigr)$ given $X^*=x$ can be tabulated as in \cref{tab:pmf} under \cref{ass:MTR}.

 \begin{table}[htbp]
\caption{The distribution of $\bigl( Y^*(0), Y^*(1), T^* \bigr)$ given $X^*=x$ under the MTR}\label{tab:pmf}
	\center  
\begin{tabular}{c|ccc|c} 
\hline\hline 
$\bigl( Y^*(0), Y^*(1) \bigr)=$  		& $(0,0)$  		&  $(0,1)$  &  	$(1,1)$ & $(1,0)$ \\   \hline
$T^*=0$     &  $q_0(x)$		&	$q_2(x)$	& 	$q_4(x)$ &  $0$ \\
$T^*=1$  	&  $q_1(x)$  	& 	$q_3(x)$	& 	$q_5(x)$ &  $0$\\ \hline
 Prob Restrictions & \multicolumn{4}{c}{$\sum_{j=0}^5 q_j(x) = 1$}   \\
			       & \multicolumn{4}{c}{$0\leq q_j(x)\leq 1$ for $j=0,\dots,5$}  \\ \hline
\end{tabular}
\end{table}

Further, \cref{ass:MTS} imposes additional restrictions on the $q_j$ functions such that 
\begin{equation}\label{eq:MTS_qj}
	\begin{aligned} 
	\frac{q_5(x)}{q_1(x)+q_3(x)+q_5(x)} &\geq \frac{q_4(x)}{q_0(x)+q_2(x)+q_4(x)}, \\
	\frac{q_3(x) + q_5(x)}{q_1(x)+q_3(x)+q_5(x)} &\geq \frac{q_2(x) + q_4(x)}{q_0(x)+q_2(x)+q_4(x)}.
	\end{aligned}
\end{equation}
Now, $q_j(x)$'s are all otherwise unrestricted under \cref{ass:MTR,ass:MTS}.

\begin{lemma} \label{lem:lower_and_upper_bounds}
Suppose that \cref{ass:support,ass:overlap,ass:MTR,ass:MTS} hold. Suppose that the distribution of $(Y^*,T^*,X^*)$ is known. For all $(t,x) \in \{0,1\}\times\mathcal{X}$,
    \begin{align*}
        (-1)^t \bigl[ \Pr\{ Y^*(t)=1\mid X^*=x \} - \Pr( Y^*=1\mid X^*=x) \bigr]\leq 0,\\
        (-1)^t \bigl[ \Pr\{ Y^*(t)=1 \mid X^*=x  \} - \Pr(Y^*=1 \mid T^*=t,X^*=x) \bigr]\geq 0,
    \end{align*}
and the bounds are sharp. 
\proof  
\Cref{ass:overlap} ensures that $0 < \Pr(T^*=t\mid X^*=x) < 1$, and hence $\Pr( Y^*=1\mid T^*=t,X^*=x)$ is well-defined. Below we will write $q_j(x)$ for $j=0,1,2,\cdots, 5$ for the probabilities that describe the distribution of $\bigl(Y^*(0),Y^*(1),T^*)$ given $X^*=x$: see our discussion in the beginning of the current section. 
    
For $(y,t)\in \{0,1\}^2$, let $\Delta_{yt}(x) := \Pr(Y^* = y, T^* = t\mid X^* = x)$. When the distribution of $(Y^*,T^*)$ given $X^*=x$ is given, we know that under \cref{ass:MTR},
    \begin{align*}
        q_0(x) + q_2(x) &= \Delta_{00}(x), \\
        q_4(x) &= \Delta_{10}(x),\\
        q_1(x) &= \Delta_{01}(x), \\
        q_3(x) + q_5(x) &= 1 - \Delta_{00}(x) - \Delta_{10}(x) - \Delta_{01}(x),
    \end{align*}
where $\sum_{j=0}^5 q_j(x) = 1$. Therefore, two of the $q_j(x)$'s are undetermined under \cref{ass:MTR}. Specifically, since $q_5(x) = 1-\sum_{j=0}^4 q_j(x)$, dropping the last equation for its redundancy yields  
    \begin{equation}\label{eq:qjs}  
    q_1(x) = \Delta_{01}(x),\quad 
    q_2(x) = \Delta_{00}(x) - q_0(x),\quad
    q_4(x) = \Delta_{10}(x),
    \end{equation}
and $q_5(x) = \Delta_{11}(x) - q_3(x)$: i.e., we have used $q_0(x)$ and $q_3(x)$ as the undetermined ones here. Since all the $q_j(x)$'s are between $0$ and $1$, $q_0(x)$ and $q_3(x)$ must satisfy $0\leq q_0(x)\leq \Delta_{00}(x)$ and $0\leq q_3(x) \leq \Delta_{11}(x)$: they are otherwise unrestricted under \cref{ass:MTR}. 
    
Now, \cref{ass:MTS} imposes additional restrictions on $q_0(x), q_3(x)$: i.e., the inequalities in \cref{eq:MTS_qj} can now be written as 
    \begin{equation}
    \begin{aligned} 
    \frac{\Delta_{11}(x) - q_3(x)}{\Delta_{01}(x)+\Delta_{11}(x)} &\geq \frac{\Delta_{10}(x)}{\Delta_{00}(x)+\Delta_{10}(x)},\\
    \frac{\Delta_{11}(x)}{\Delta_{01}(x) + \Delta_{11}(x)} &\geq \frac{\Delta_{00}(x) + \Delta_{10}(x) - q_0(x)}{\Delta_{00}(x) + \Delta_{10}(x)}.
    \end{aligned}
    \end{equation}
Therefore, under \cref{ass:MTR,ass:MTS}, we have 
    \begin{align} 
    0\leq q_3(x) \leq \Delta_{11}(x) - \frac{\Delta_{10}(x)\{\Delta_{01}(x) + \Delta_{11}(x)\}}{\Delta_{00}(x) + \Delta_{10}(x)},  \label{eq:q3}\\
    \Delta_{00}(x)+\Delta_{10}(x) - \frac{\Delta_{11}(x)\{\Delta_{00}(x)+\Delta_{10}(x) \}}{\Delta_{01}(x)+\Delta_{11}(x)}   \leq q_0(x) \leq \Delta_{00}(x). \label{eq:q0}
    \end{align}
These are the only restrictions that \cref{ass:MTR,ass:MTS} impose on $q_0(x)$ and $q_3(0)$ when we know the distribution of $(Y^*,T^*)$ given $X^*=x$. Note that the upper bound in \cref{eq:q3} is equal to 
    \begin{align*} 
    &\Pr(Y^*=1, T^*=1\mid X^*=x) - \Pr(T^*=1\mid X^*=x)\Pr(Y^*=1\mid T^* = 0,X^*=x)  \\
    &=
    \Pr(T^*=1\mid X^*=x)\bigl\{ \Pr(Y^*=1\mid T^*=1,X^*=x) - \Pr(Y^*=1\mid T^*=0,X^*=x) \bigr\},
    \end{align*} 
which is trivially no greater than $1$ and is no smaller than $0$ by \cref{ass:MTR,ass:MTS}, because 
    \begin{equation} \label{eq:explanation}
    \begin{aligned}  
        &\Pr(Y^*=1\mid T^*=1,X^*=x) - \Pr(Y^*=1\mid T^*=0,X^*=x) \\
        &=
        \Pr\{Y^*(1)=1\mid T^*=1,X^*=x\} - \Pr\{Y^*(0)=1\mid T^*=0,X^*=x\} \\
        &\geq 
        \Pr\{Y^*(1)=1\mid T^*=1,X^*=x\} - \Pr\{Y^*(1)=1\mid T^*=0,X^*=x\}
        \geq 
        0.
    \end{aligned}
    \end{equation}
The lower bound in \cref{eq:q0} is equal to 
    \begin{equation*}
    \Pr(T^*=0\mid X^*=x) - \Pr(T^*=0\mid X^*=x)\Pr(Y^*=1\mid T^*=1,X^*=x),
    \end{equation*}
which is trivially no smaller than $0$ and is no greater than $\Delta_{00}(x)$ by \cref{ass:MTR,ass:MTS}, because 
    \begin{align*} 
        1 - \Pr(Y^*=1\mid T^*=1,X^*=x)
        \leq 
        1 - \Pr(Y^*=1\mid T^*=0,X^*=x)
    %	=
    %	\Pr(Y^*=0\mid T^*=0,X^*=x)
    \end{align*} 
as we have shown in \cref{eq:explanation}. 
    
Below we calculate the sharp bounds on $\Pr\{ Y^*(t)=1\mid X^*=x\}$ for $t=0,1$ by expressing them as a function of $q_0(x), q_3(x)$ and finding their maxima and minima subject to the constraints in \cref{eq:q3,eq:q0}. First, consider the case of $t=1$:  
    \begin{align*} 
    \Pr\{ Y^*(1)=1\mid X^* = x\}
    &=
    \Pr\{ Y^*(1)=1, T^*=0\mid X^* = x\} + \Delta_{11}(x) \\
    &=
    q_2(x) + q_4(x) + \Delta_{11}(x)\\
    &=
    \Delta_{00}(x) + \Delta_{10}(x) + \Delta_{11}(x) - q_0(x).
    \end{align*}
Therefore, by using \cref{eq:q0}, we can find the sharp bounds on $\Pr\{ Y^*(1)=1\mid X^* = x\}$. Specifically, when $q_0(x) = \Delta_{00}(x)$, we have the sharp lower bound on $\Pr\{ Y^*(1)=1\mid X^* = x\}$, which is equal to
    \[ 
        \Delta_{00}(x) + \Delta_{10}(x) + \Delta_{11}(x) - \Delta_{00}(x)
        = \Pr(Y^*=1\mid X^*=x).
    \]
    
 For the sharp upper bound on $\Pr\{ Y^*(1)=1\mid X^* = x\}$, we set $q_0(x)$ to its lower bound in \cref{eq:q0}, which yields 
    \begin{align*}
        &\Delta_{00}(x) + \Delta_{10}(x) + \Delta_{11}(x) - \Delta_{00}(x)-\Delta_{10}(x) + \frac{\Delta_{11}(x)\{\Delta_{00}(x)+\Delta_{10}(x) \}}{\Delta_{01}(x)+\Delta_{11}(x)} \\
        &=
        \Delta_{11}(x)\Bigl\{ 1+ \frac{\Delta_{00}(x)+\Delta_{10}(x)}{\Delta_{01}(x)+\Delta_{11}(x)}\Bigr\} \\
        &=
        \Pr(Y^*=1,T^*=1\mid X^*=x)
        \Bigl\{ 1 + \frac{\Pr(T^*=0\mid X^*=x)}{\Pr(T^*=1\mid X^*=x)} \Bigr\} \\
        &= 
        \frac{\Pr(Y^*=1,T^*=1\mid X^*=x)}{\Pr(T^*=1\mid X^*=x)}.
    \end{align*}
These upper and lower bounds are sharp: indeed, $\Pr\{ Y^*(1)=1\mid X^* = x\}$ can attain any value $\delta_1(x)$ in the described interval by setting $q_0(x) = \Delta_{00}(x)+\Delta_{10}(x) +\Delta_{11}(x) - \delta_1(x)$.
    
The case of $t=0$ is similar: i.e., note that  
    \[
        \Pr\{ Y^*(0)=1\mid X^* = x\}
        =
        \Delta_{10}(x) + \Delta_{11}(x) - q_3(x),
    \]
and use the bounds on $q_3(x)$ in \cref{eq:q3}: $\Pr\{ Y^*(0) = 1\mid X^*=x)$ can attain any value $\delta_0(x)$ in the described interval by setting $q_3(x) = \Delta_{10}(x) + \Delta_{11}(x) - \delta_0(x)$. 
    
Once we choose the values of $q_0(x)$ and $q_3(x)$ such that $\Pr\{ Y^*(t) = 1\mid X^*=x) = \delta_t(x)$ for $t=0,1$, where $\delta_t(x)$'s are arbitrary points in the stated intervals on $\Pr\{Y^*(t) =1\mid X^*=x\}$, we can now determine all the other $q_j(x)$'s by \cref{eq:qjs}.     \qed
\end{lemma}

\noindent
\textbf{Remark: } 
The proof of \cref{lem:lower_and_upper_bounds} shows that \cref{ass:MTS} is not needed for the sharp lower bound on $\Pr\{ Y^*(1)=1\mid X^*=x)$ and the sharp upper bound on $\Pr\{ Y^*(0) = 1\mid X^*=x \}$. 
In other words, the first displayed line in the conclusion of \cref{lem:lower_and_upper_bounds} does not hinge on \cref{ass:MTS}.

\begin{lemma}\label{lem:r0}
	Suppose that \cref{ass:support} holds. For $d\in \{\CC,\CP\}$ and for all $x\in \mathcal{X}$, we have $r_d(x, p_0) = \Pr(Y^* = 1\mid X^*=x)$. 
	\proof
	We focus on \cref{design1}, i.e., the case of $d = \CC$; \cref{design2} is similar but simpler. By the Bayes rule and the sampling design,
	\begin{multline}\label{eq:YsXs}
	\Pr(Y^*=1\mid X^*=x)
	=
	\frac{p_0 f_{X^*|Y^*}(x\mid 1)}{p_0f_{X^*|Y^*}(x\mid 1) +(1-p_0) f_{X^*|Y^*}(x\mid 0)}
	\\
	=
	\frac{p_0 f_{X|Y}(x\mid 1)}{p_0f_{X|Y}(x\mid 1) +(1-p_0) f_{X|Y}(x\mid 0)}.
	\end{multline}
	Here, by the Bayes rule again, for $y=0,1$,
	\begin{equation}\label{eq:XY}
	f_{X|Y}(x|y)
	=
	\frac{f_X(x) \Pr(Y=y\mid X=x)}{\Pr(Y=y)}.
	\end{equation}
	Combining \cref{eq:YsXs,eq:XY} yields the result. \qed
\end{lemma}

\begin{lemma}\label{lem:bayes}
	Suppose that \cref{ass:support,ass:overlap} are satisfied. Then, for all $(t,x) \in \{0,1\}\times \mathcal{X}$, 
	\begin{multline*}
	\Pr(Y^*=1 \mid T^*=t,X^*=x)	 \\
	= 
	\left\{
	\begin{aligned}
	&\frac{r_\CC(x,p_0)\Pi(t\mid 1,x)}{\Pi(t\mid 0,x) + r_\CC\{x,p_0\}\{\Pi(t\mid 1,x) - \Pi(t\mid 0,x) \}}, && \text{under \cref{design1},} \\
	&\frac{r_\CP(x,p_0)\Pi(t\mid 1,x)}{\Pi(t\mid 0,x) }, &&\text{under \cref{design2}.}
	\end{aligned}
	\right.
	\end{multline*}
	\proof
	Under \cref{design1}, $\Pi(t\mid y,x)$ is equal to $\Pr(T^*=t\mid Y^*=y, X^*=x)$, which is contained in the interval $(0,1)$ by \cref{ass:overlap}. Therefore, the asserted formula is well-defined, and the result follows from \cref{lem:r0} and the Bayes rule. Similarly, under \cref{design2}, $\Pi(t\mid 1,x)$ is equal to $\Pr(T^*=t\mid Y^*=1,X^*x=x)$ while $\Pi(t\mid 0,x)$ is equal to $\Pr(T^*=t\mid X^*=x)$, where they are all in $(0,1)$ by \cref{ass:overlap}. Therefore, the stated formula is well-defined, and the conclusion follows from \cref{lem:r0} and the Bayes rule. 
	\qed
\end{lemma}

\begin{lemma}\label{lem:ordrr}
	Suppose that \cref{ass:support,ass:overlap} are satisfied. Then, for all $x\in \mathcal{X}$, 
	\[
	\frac{\Pr(Y^*=1\mid T^*=1,X^*=x)}{\Pr(Y^* = 1 \mid T^*=0,X^*=x)}
	=
	\left\{ 
	\begin{aligned}
		&\Gamma_{\CC,\RR}(x, p_0) &&\text{under \cref{design1},} \\
		&\OR(x) && \text{under \cref{design2}.}
	\end{aligned}
	\right. 
	\]
	\proof
	It directly follows from \cref{lem:bayes}. \qed
\end{lemma}

\begin{lemma}\label{lem:ineq p0}
	Suppose that \cref{ass:support,ass:overlap,ass:MTR,ass:MTS} hold. Then, under \cref{design1}, for all $x\in \mathcal{X}$, we have $\Gamma_{\CC,\RR}(x,p_0)\leq \Gamma_{\CC,\RR}(x,0)$. 
	\proof
	Since $\Gamma_{\CC,\RR}(x,0) = \OR(x)$, it follows from the invariance property of the odds ratio and \cref{lem:ordrr} that 
	\begin{equation*}
	\Gamma_{\CC,\RR}(x,0) = \Gamma_{\CC,\RR}(x,p_0)\frac{\Pr(Y^*=0\mid T^*=0,X^*=x)}{\Pr(Y^*=0\mid T^*=1,X^*=x)}.
	\end{equation*}
	So, it suffices to show $\Pr(Y^*=1\mid T^*=1,X^*=x)\geq \Pr(Y^*=1\mid T^*=0,X^*=x)$, which directly follows from \cref{ass:MTR,ass:MTS}. \qed
\end{lemma}

\subsection{Proofs of the results in section \ref{sec:eff:influence}}
In this subsection, we use the following notation: for a function $g(\cdot\ ; \gamma)$ that depends on $\gamma$, we will write $\partial_\gamma g(x; \gamma_0)$ for $\partial g(x; \gamma)/\partial \gamma \big|_{\gamma = \gamma_0}$.

\noindent\textbf{Proof of \Cref{lem:tangent}: } 
First, the likelihood of a single observation $(Y,T,X)$ is given by
\begin{equation}\label{eq:likelihood}
L(Y,T,X) = \bigl\{ (1-h_0) \mathcal{P}_0(T,X) \bigr\}^{1-Y} \bigl\{ h_0 \mathcal{P}_1(T,X) \bigr\}^Y,
\end{equation}
where $\mathcal{P}_y(T,X) = f_{X|Y}(X\mid y) \Pr(T=1 \mid X,Y=y)^T \bigl\{ 1- \Pr(T=1\mid X,Y=y) \bigr\}^{1-T}$ for $y=0,1$.  Let $\gamma$ be the parameter denoting regular parametric submodels, where the true value will be denoted by $\gamma_0$.  Then, the score evaluated at $\gamma_0$ is equal to 
\begin{multline}\label{eq:scores}
(1-Y) 
\Bigl[
S_{X|Y}(X\mid 0)  
+
\frac{\bigl\{ T-\Pr(T=1\mid X,Y=0) \bigr\} \partial_\gamma \Pr(T=1\mid X,Y=0; \gamma_0)}{\Pr(T=1\mid X,Y=0) \{ 1- \Pr(T=1\mid X,Y=0)\}}
\Bigr] \\
+
Y 
\Bigl[
S_{X|Y}(X\mid 1)
+
\frac{ \bigl\{ T-\Pr(T=1\mid X,Y=1) \bigr\} \partial_\gamma \Pr(T=1\mid X,Y=1; \gamma_0)}{\Pr(T=1\mid X,Y=1) \{ 1- \Pr(T=1\mid X,Y=1)\}}
\Bigr],
\end{multline}
where $S_{X|Y}(x\mid y) = \partial_\gamma \log f_{X|Y}(x\mid y; \gamma_0)$ is restricted only by $\Exp\{ S_{X|Y}(X\mid y) \mid Y=y \} = 0$, while the derivatives $\partial_\gamma \Pr(T=1\mid X,Y=y,\gamma_0)$ are unrestricted. \qed

\noindent\textbf{Proof of \cref{thm:eff}: } For brevity, we focus on $\beta:= \beta(0)$: a proof for $\beta(1)$ is analogous.  Let $p_0(x) := \Pr(T=1\mid X=x,Y=0)$ and $p_1(x) := \Pr(T=1 \mid X=x,Y=1)$, and we have
\begin{equation*}
\beta(\gamma)
:=
\int_{\mathcal{X}} \log \Biggl[ \, \underbrace{\frac{p_1(x; \gamma)}{1-p_1(x; \gamma)} \cdot \frac{1-p_0(x;\gamma)}{p_0(x; \gamma)}}_{:= \OR(x;\gamma)} \, \Biggr] f_{X|Y}(x\mid 0; \gamma) dx,
\end{equation*}
where $\gamma$ represents regular parametric submodels such that $\gamma_0$ is the truth. Since
\begin{equation}\label{eq:dOR}
\partial_\gamma\OR(x;\gamma_0)
=
%\partial_\gamma p_1(x;\gamma_0) \frac{\{ 1-p_0(x)\}}{p_0(x)\{ 1-p_1(x)\}^2}
%- 
%\partial_\gamma p_0(x;\gamma_0) \frac{p_1(x)}{p_0^2(x)\{ 1-p_1(x)\}} \\
%=
\underbrace{\frac{\partial_\gamma p_1(x;\gamma_0)}{p_1(x)\{1-p_1(x)\}}}_{=: A_1(x)} \OR(x)
-
\underbrace{\frac{\partial_\gamma p_0(x;\gamma_0)}{p_0(x)\{ 1-p_0(x)\}}}_{=: A_0(x)} \OR(x),
\end{equation}
we obtain
\begin{equation}\label{eq:dbeta}
\partial_\gamma \beta(\gamma_0)
%=
%\int \Bigl[ \frac{\partial_\gamma\OR(x;\gamma_0)}{\OR(x)} + \log \{ \OR(x) \} S_{X|Y}(x|0) %\Bigr] f_{X|Y}(x|0) dx \\
=
\int \bigl[ 
%\frac{\partial_\gamma p_1(x;\gamma_0)}{p_1(x)\{1-p_1(x)\}}
%-
%\frac{\partial_\gamma p_0(x;\gamma_0)}{p_0(x)\{ 1-p_0(x)\}}
A_1(x) 
- 
A_0(x)
+
\log \{ \OR(x) \} S_{X|Y}(x\mid 0) 
\bigr] f_{X|Y}(x\mid 0) dx.
\end{equation}
Now, we only need to verify the equality between $\Exp\{ F_0(Y,T,X) S(Y,T,X)  \}$ and the right-hand side of \cref{eq:dbeta}, where $F_0(Y,T,X)$ and $S(Y,T,X)$ are given in the theorem statement and \eqref{eq:scores}, respectively.  
%\begin{equation*}
%\begin{aligned}
%&S(Y,T,X) \\
%&=
%(1-Y) 
%\Bigl[
%S_{X|Y}(X|0)  
%+
%\bigl\{ T-p_0(X) \bigr\} A_0(X) 
%\Bigr] 
%+
%Y 
%\Bigl[
%S_{X|Y}(X|1)
%+
%\bigl\{ T-p_1(X) \bigr\} A_1(X)
%\Bigr], \\
%&F_0(Y,T,X) \\
%&=
%\frac{1-Y}{1-h_0}
%\Biggl[
%\log \OR(X) - \beta
%-
%\frac{\bigl\{ T-p_0(X)  \bigr\}}{p_0(X) \{  1- p_0(X)\} }
%\Biggr] 
%+ 
%\frac{Y}{h_0}  \frac{f_{X|Y}(X|0)}{f_{X|Y}(X|1)}	
%\frac{ \bigl\{ T-p_1(X)  \bigr\}}{p_1(X)\{1-p_1(X) \}}.
%\end{aligned}
%\end{equation*}
Note that $F_0(Y,T,X) S(Y,T,X)$ is equal to
\begin{multline*}
\frac{1-Y}{1-h_0}
\Biggl[
\log \OR(X) - \beta
-
\frac{ \bigl\{ T-p_0(X)  \bigr\}}{p_0(X) \{  1- p_0(X)\} }
\Biggr] \Bigl[
S_{X|Y}(X\mid 0)  
+
\bigl\{ T-p_0(X) \bigr\} A_0(X) 
\Bigr] \\
+ 
\frac{Y}{h_0} \frac{f_{X|Y}(X\mid 0)}{f_{X|Y}(X\mid 1)}
\Biggl[
\frac{ \bigl\{ T-p_1(X)  \bigr\}}{p_1(X)\{1-p_1(X) \}}
\Biggr]
\Bigl[
S_{X|Y}(X\mid 1)
+
\bigl\{ T-p_1(X) \bigr\} A_1(X)
\Bigr].
\end{multline*}
Here, taking expectations directly shows that $\Exp\bigl\{ F_0(Y,T,X) S(Y,T,X)  \bigr\}$ is equal to
\begin{equation*}
\Exp\bigl[ \log \{\OR(X) \} S_{X|Y}(X|0) -  A_0(X) \ \big| \ Y=0 \bigr] 
+  
\Exp\biggl\{ \frac{f_{X|Y}(X|0)}{f_{X|Y}(X|1)}  A_1(X)  \ \bigg| \ Y=1 \biggr\},
\end{equation*}
which is equal to the right-hand side of \eqref{eq:dbeta} because 
\begin{equation*}
\Exp\biggl\{ \frac{f_{X|Y}(X|0)}{f_{X|Y}(X|1)}  A_1(X)  \ \bigg| \ Y=1 \biggr\}
= 
\Exp\bigl\{  A_1(X) \ \big| \ Y=0 \bigr\}. 
\end{equation*}
Finally, it follows from \cref{lem:tangent} that $F_0$ is an element of the tangent space. \qed

\subsection{Proofs of the results in the main text}\label{appx:maintextproofs}

\subsubsection{Proofs of the results for causal relative risk} 

\noindent \textbf{Proof of \cref{lem:rr benchmark}: } If \cref{ass:support,ass:overlap,ass:unconfounding} are satisfied, the independence of $T^*$ and $Y^*(t)$ given $X^*=x$ yields 
\begin{equation*}
    \frac{\Pr(Y^*=1\mid T^*=1,X^*=x)}{\Pr(Y^*=1\mid T^*=0,X^*=x)}
    =
    \frac{\Pr\{Y^*(1)=1\mid X^*=x\}}{\Pr\{Y^*(0)=1\mid X^*=x\}} 
    =
    \theta_\RR(x),
\end{equation*}
where $\Pr(Y^*=1\mid T^*=0,X^*=x) = \Pr\{Y^*(0)=1\mid X^*=x\} >0 $ by \cref{ass:overlap,ass:unconfounding}.

Alternatively, suppose that \cref{ass:support,ass:overlap,ass:MTR,ass:MTS} are satisfied. Then, \cref{ass:overlap} ensures that $\Pr\{ Y^*(0) =1 \mid X^* = x\} >0$. Therefore, the sharp lower bound on $\theta_\RR(x)$ under \cref{ass:support,ass:overlap,ass:MTR} follows from \cref{lem:lower_and_upper_bounds}: i.e., $\Pr(Y^*(1)=1\mid X^*=x) \geq \Pr(Y^*=1\mid X^*=x)$ and $\Pr(Y^*(0)=1\mid X^*=x) \leq \Pr(Y^*=1\mid X^*=x)$. Further, from \cref{lem:lower_and_upper_bounds}, we have a sharp upper bound on $\theta_\RR(x)$ under \cref{ass:MTS} such that
\begin{equation}\label{eq:first sharp}
\theta_\RR(x) \leq \frac{\Pr(Y^*=1\mid T^*=1,X^*=x)}{\Pr(Y^*=1\mid T^*=0,X^*=x)}.  \qedhere
\end{equation}

\noindent \textbf{Proof of \cref{thm:unconfounding}: } 
First, consider the case of \cref{design1}. For $(y,t,x) \in \{0,1\}^2\times \mathcal{X}$, we have 
\[
\Pi(t\mid y,x) 
=
\Pr(T= t\mid Y=y, X=x)
=
\Pr(T^*= t\mid Y^*=y, X^*=x). 	
\]
Therefore, using the fact that $r_\CC(x,p_0) = \Pr(Y^*=1\mid X^*=x)$ by \cref{lem:bayes}, we obtain 
\begin{align*}
\Gamma_{\CC,\RR}(x,p_0)
=
\frac{\Pr(T^*=1\mid Y^*=1,X^*=x)}{\Pr(T^*=0\mid Y^*=1,X^*=x)}
\frac{\Pr(T^*=0\mid X^*=x)}{\Pr(T^*=1\mid X^*=x)}
\end{align*}
Multiplying $\Pr(Y^*=1\mid X^*=x)$ to the numerator and the denominator of the right-hand side yields 
\[
	\Gamma_{\CC,\RR}(x,p_0)
	=
	\frac{\Pr(Y^* = 1, T^* = 1\mid X^* = x)\Pr(T^* = 0\mid X^* = x)}{\Pr(Y^* = 1, T^* = 0\mid X^* = x)\Pr(T^* = 1\mid X^*=x)}.	
\]
Therefore, the conclusion follows from \cref{ass:unconfounding}. Now, consider the case of \cref{design2}. Then, for $(t,x) \in \{0,1\}\times \mathcal{X}$, we have 
\[
\Pi(t\mid 1,x) = \Pr(T^*=t\mid Y^*=1,X^*=x)
\quad \text{and}\quad
\Pi(t\mid 0,x) = \Pr(T^*=t\mid X^*=x). 	
\]
Therefore, using the definition of $\Gamma_{\CP,\RR}(x,p)$ and the fact that $r_\CP(x,0) = 0$, we obtain
\[
\Gamma_{\CP,\RR}(x,0) 
= 
\frac{\Pr(T^*=1\mid Y^*=1,X^* = x)}{\Pr(T^*=0\mid Y^*=1,X^* = x)}	
\frac{\Pr(T^* = 0\mid X^* = x)}{\Pr(T^* = 1\mid X^* = x)}.
\]
Now, multiplying $\Pr(Y^*=1\mid X^*=x)$ to the numerator and the denominator of the right-hand side and using \cref{ass:unconfounding} proves the claim. \qed

\noindent \textbf{Proof of \cref{cor:unconfounding}: }  We show that $\Gamma_{\CC,\RR}(x,p)$ is monotonic in $p$. Since $\partial r_\CC(x,p) /\partial p > 0$ under \cref{ass:overlap}, it suffices to consider the derivative of $\tilde \Gamma_x(r)$, where 
\[
    \tilde \Gamma_x(r) := \frac{ a_x + r(b_x-a_x) }{ 1-a_x + r(a_x-b_x) },
\] 
where $a_x:= \Pi(0\mid 0,x)$ and $b_x:=\Pi(0\mid 1,x)$. But, direct calculation shows that
\begin{equation*}
\frac{\partial \tilde \Gamma_x(r)}{\partial r}  
=
\frac{b_x-a_x}{\bigl\{ 1-a_x + r(a_x-b_x) \bigr\}^2},
\end{equation*}
which does not change the sign as $r$ varies. Therefore, $\Gamma_{\CC,\RR}(x,p)$ is either increasing or decreasing in $p$: i.e.,
\begin{align*}
\min \{ \Gamma_{\CC,\RR}(x,0), \Gamma_{\CC,\RR}(x,\bar p)\} 
\leq   
\Gamma_{\CC,\RR}(x,p_0) 
\leq 
\max \{ \Gamma_{\CC,\RR}(x,0), \Gamma_{\CC,\RR}(x,\bar p)\} 
\end{align*}
Now, note that $\Gamma_{\CC,\RR}(x,0) = \OR(x)$, and use \cref{thm:unconfounding}. \qed

\noindent
\textbf{Remark: } Continue to consider \cref{design1} and assume that \cref{ass:support,ass:overlap,ass:unconfounding} are satisfied.  Using the notation used in the proof of \cref{cor:unconfounding}, we have  
\begin{align*}
b_x - a_x
&=
\Pr(T^*=0\mid X^*=x)\left\{ \frac{\Pr\{Y^*(0)=1\mid X^*=x\}}{\Pr(Y^*=1\mid X^*=x)} - \frac{\Pr\{Y^*(0)=0\mid X^*=x\}}{\Pr(Y^*=0\mid X^*=x)}  \right\}
\\
&=
\frac{\Pr(T^*=0\mid X^*=x)\Bigl[ \Pr\{Y^*(0)=1\mid X^*=x\} - \Pr(Y^*=1\mid X^*=x)  \Bigr]}{\Pr(Y^*=1\mid X^*=x)\Pr(Y^*=0\mid X^*=x)},
\end{align*}
where
\begin{multline*}
\Pr\{Y^*(0)=1\mid X^*=x\} - \Pr(Y^*=1\mid X^*=x) \\
=
\Pr\{Y^*(0)=1, T^*=1\mid X^*=x\}  
-
\Pr\{Y^*(1)=1, T^*=1\mid X^*=x\}.
\end{multline*}
Therefore, if $Y^*(1)\geq Y^*(0)$ almost surely (i.e., \cref{ass:MTR}), then we must have $b_x - a_x\leq 0$, which means that $\Gamma_{\CC,\RR}(x,p)$ is decreasing in $p$.  

\noindent\textbf{Proof of \cref{thm:bounds}: } 
From \cref{lem:rr benchmark}, we obtain under \cref{ass:support,ass:overlap,ass:MTR,ass:MTS} that 
\[
   1\leq  \theta_\RR(x) \leq \frac{\Pr(Y^*=1\mid T^*=1,X^*=x)}{\Pr(Y^*=1\mid T^*=0,X^*=x)},
\]
where the bounds are sharp. Therefore, by \cref{lem:ordrr}, the upper bound becomes 
\[
	\theta_\RR(x) \leq 
    \left\{ 
        \begin{aligned}
            &\Gamma_{\CC,\RR}(x,p_0)  &&\text{under \cref{design1},}\\
            &\OR(x)  &&\text{under \cref{design2},}
        \end{aligned}
    \right.
\]
which would be sharp if $p_0$ were given. We are done for the case of \cref{design2} because the bound does not depend on $p_0$. In the case of \cref{design1}, we use \cref{lem:ineq p0} and sharpness follows from the continuity of $\Gamma_{\CC,\RR}(x,\cdot)$.   \qed

\subsubsection{Proofs of the results for causal attributable risk} 

\noindent\textbf{Proof of \cref{lem:ar benchmark}: }
If \cref{ass:support,ass:overlap,ass:unconfounding} are satisfied, then the independence of $T^*$ and $Y^*(t)$ given $X^*=x$ yields 
\begin{multline*}
    \Pr( Y^*=1\mid T^*=1,X^*=x) - \Pr( Y^*=1\mid T^*=0,X^*=x)
    \\
    =
    \Pr\{  Y^*(1)=1\mid X^*=x\} - \Pr\{ Y^*(0)=1\mid X^*=x \} 
    =
    \theta_\AR(x).
\end{multline*}
Alternatively, suppose that \cref{ass:support,ass:overlap,ass:MTR,ass:MTS} are satisfied. Then, \cref{ass:MTR} leads to the sharp lower bound of $0$ on $\theta_\AR(x)$ by \cref{lem:lower_and_upper_bounds}. Similarly, the sharp upper bound on $\theta_\AR(x)$ under \cref{ass:MTS} follows from \cref{lem:lower_and_upper_bounds}. \qed

\noindent\textbf{Proof of \cref{thm:unconfounding AR}: } 
First, consider the case of \cref{design1}. For $(y,t,x) \in \{0,1\}^2\times \mathcal{X}$, we have 
\[
\Pi(t\mid y,x) 
=
\Pr(T= t\mid Y=y, X=x)
=
\Pr(T^*= t\mid Y^*=y, X^*=x),	
\]
from which $\Gamma_{\CC,\AR}(x,p)$ can be written as 
\begin{align*}
&
\frac{\Pr(T^*=1\mid Y^*=1, X^*=x)}{r_\CC(x,p)\Pr(T^*=1\mid Y^*=1, X^*=x) + \{1-r_\CC(x,p)\}\Pr(T^*=1\mid Y^*=0, X^*=x) \}}
\\
& 
-
\frac{\Pr(T^*=0\mid Y^*=1, X^*=x)}{r_\CC(x,p)\Pr(T^*=0\mid Y^*=1, X^*=x) + \{1-r_\CC(x,p)\}\Pr(T^*=0\mid Y^*=0, X^*=x) \}}.
\end{align*}
Therefore, by \cref{lem:bayes}, 
\begin{equation}\label{eq:ar-cc-equality}
r_\CC(x,p_0) \Gamma_{\CC,\AR}(x,p_0)
=
\frac{\Pr(Y^*=1,T^*=1\mid X^*=x)}{\Pr(T^*=1\mid X^*=x)}	
-
\frac{\Pr(Y^*=1,T^*=0\mid X^*=x)}{\Pr(T^*=0\mid X^*=x)}.	
\end{equation}
Therefore, the conclusion follows from \cref{ass:unconfounding}.

Now, consider the case of \cref{design2}. Then, for $(t,x) \in \{0,1\}\times \mathcal{X}$, we have 
\[
\Pi(t\mid 1,x) = \Pr(T^*=t\mid Y^*=1,X^*=x)
\quad \text{and}\quad
\Pi(t\mid 0,x) = \Pr(T^*=t\mid X^*=x). 	
\]
Hence, $\Gamma_{\CP,\AR}(x,0)$ is given by 
\begin{equation*}
	\frac{\Pr(T^*=1\mid Y^*=1, X^*=x)}{\Pr(T^*=1\mid X^*=x) }
	-
	\frac{\Pr(T^*=0\mid Y^*=1, X^*=x)}{\Pr(T^*=0\mid X^*=x) },
\end{equation*}
because $r_\CP(x,0) = 0$. Therefore, by \cref{lem:bayes}, 
\begin{equation}\label{eq:ar-cp-equality}
r_\CP(x,p_0) \Gamma_{\CP,\AR}(x,0)
=
\frac{\Pr(Y^*=1, T^*=1\mid X^*=x)}{\Pr(T^*=1\mid X^*=x) } 
-
\frac{\Pr(Y^*=1,T^* = 0\mid X^*=x)}{\Pr(T^*=0\mid X^*=x) }.
\end{equation}
So, the conclusion follows from \cref{ass:unconfounding}. \qed

\noindent\textbf{Proof of \cref{cor:unconfounding AR}: }  In the case of \cref{design1}, we first recall from \cref{thm:unconfounding AR} that $\theta_\AR(x) = r_\CC(x,p_0) \Gamma_{\CC,\AR}(x,p_0)$, where $p_0$ is unidentified. Since $r_\CC(x,p) \Gamma_{\CC,\AR}(x,p)$ is continuous in $p$, the sharp bounds on $\theta_\AR(x)$ can be obtained by taking maximum and minimum over $p\in[0,\bar p]$ under \cref{ass:rare}. Here, we note that $\theta_\AR(x) \in [-1,1]$ by definition, but this information does not provide anything extra, because $r_\CC(x,p) \Gamma_{\CC}(x,p) \in [-1,1]$ for all $x,p$ by definition.  To see this point, just note that $r_\CC(x,p) \Gamma_{\CC}(x,p)$ has the form of 
\[
\frac{ra_x}{ra_x + (1-r)b_x} - \frac{r(1-a_x)}{r(1-a_x) + (1-r)(1-b_x)},	
\]
where $r, a_x, b_x$ are all between $0$ and $1$. For the case of \cref{design2}, we recall from \cref{thm:unconfounding AR} that $\theta_\AR(x) = r_\CP(x,p_0) \Gamma_{\CP,\AR}(x,0)$, where $p_0$ is unidentified. Since $r_\CP(x,p) \Gamma_{\CP,\AR}(x,0)$ is continuous and monotonic in $p$ with $r_\CP(x,0) = 0$, the sharp bounds on $\theta_\AR(x)$ in this case will be either $[0,\ r_\CP(x,\bar p) \Gamma_{\CP,\AR}(x,0)]$ or $[r_\CP(x,\bar p) \Gamma_{\CP,\AR}(x,0),\ 0]$, depending on the sign of $\Gamma_{\CP,\AR}(x,0)$. Similarly to the case of \cref{design1}, the fact that $\theta_\AR(x)\in[-1,1]$ does not provide anything extra, because $r_\CP(x, p) \Gamma_{\CP,\AR}(x,0)$ is always between $1$ and $-1$ for all $x$ and $p\leq \bar p^*$, where $\bar p^*$ is defined in \cref{eq:pbarstar}. To see this point, note that $r_\CP(x, p) \Gamma_{\CP,\AR}(x,0)$ is equal to 
\begin{align} 
&
\frac{p(1-h_0)}{h_0}\frac{\Pr(Y=1\mid X=x)}{\Pr(Y=0\mid X=x)}
\Biggl\{  \frac{\Pr(T=1\mid Y=1,X=x)}{\Pr(T=1\mid Y=0,X=x)} - \frac{\Pr(T=0\mid Y=1,X=x)}{\Pr(T=0\mid Y=0,X=x)}	\Biggr\} \notag
\\
=   
&
p\Biggl\{ \frac{f_{T,X|Y}(1,x\mid 1)}{f_{T,X|Y}(1,x\mid 0)} - \frac{f_{T,X|Y}(0,x\mid 1)}{f_{T,X|Y}(0,x\mid 0)} \Biggr\},
\label{eq:ar bound design2}
\end{align}
where the equality follows from the Bayes rule. Here, by the definition of $\bar p^*$ in \cref{eq:pbarstar}, we must have 
\begin{equation}\label{eq:pbarstar2}
0 \leq p \frac{f_{T,X|Y}(t,x\mid 1)}{f_{T,X|Y}(t,x\mid 0)} \leq 1,
\end{equation}
for all $x, t$, and $p \leq \bar p^*$. Therefore, combining \cref{eq:ar bound design2} with \cref{eq:pbarstar2} shows that $r_\CP(x, p) \Gamma_{\CP,\AR}(x,0)$ is between $-1$ and $1$ for all $x$ and $p\leq \bar p^*$. \qed

\noindent\textbf{Proof of \cref{thm:bounds AR}: } 
Fix $x\in \mathcal{X}$, and recall from \cref{lem:ar benchmark} that 
\begin{equation}\label{eq:ar bounds}
0\leq \theta_\AR(x) \leq \Pr(Y^* = 1\mid T^* = 1,X^*=x) - \Pr( Y^* = 1\mid T^* = 0, X^*= x),
\end{equation}
where the bounds are sharp.  Now, consider the case of \cref{design1}. Since \cref{eq:ar-cc-equality} is valid under \cref{ass:support,ass:overlap}, we know that the expression on the utmost right-hand side of \cref{eq:ar bounds} is equal to $r_\CC(x,p_0) \Gamma_{\CC,\AR}(x,p_0)$ in this sampling scenario. Here, $p_0$ is the only unknown component, hence we obtain 
\[
	0 \leq \theta_\AR(x) \leq \max_{p\in [0,\bar p]} r_\CC(x,p) \Gamma_{\CC,\AR}(x,p)
\]
under \cref{ass:rare}, where sharpness follows from the continuity of $r_\CC(x,\cdot) \Gamma_{\CC,\AR}(x,\cdot)$.  Alternatively, in the case of \cref{design2}, we note that \cref{eq:ar-cp-equality} is valid under \cref{ass:support,ass:overlap}, and therefore the expression on the utmost right-hand side of \cref{eq:ar bounds} becomes $r_\CP(x,p_0)\Gamma_{\CP,\AR}(x,0)$. Here, $p_0$ is the only unknown component, and $r_\CP(x,p)$ is increasing in $p$. Therefore, we obtain 
\[
0\leq \theta_\AR(x)\leq r_\CP(x,\bar p) \Gamma_{\CP,\AR}(x,0)	
\]
under \cref{ass:rare}, where sharpness follows from the continuity of $r_\CP(x,\cdot) \Gamma_{\CP,\AR}(x,0)$.   \qed

\subsubsection{Proofs of the results for aggregated causal parameters} 

\noindent\textbf{Proof of \cref{cor:aggregation1}: } 
First, consider the case of \cref{design1}. In this sampling scenario, we have $f_{X^*}(\cdot) = f_{X\mid Y}(\cdot\mid 1) p_0 + f_{X\mid Y}(\cdot\mid 0)(1-p_0)$. Therefore, it follows from \cref{thm:unconfounding} that
\begin{multline*}
\bar\vartheta_\RR 
= 
p_0 \underbrace{ \int_\mathcal{X} \log\{ \Gamma_{\CC,\RR}(x,p_0) \} f_{X\mid Y}(x\mid 1) dx}_{=\Psi_{\CC,\RR}(p_0,1)} \\
+ 
(1-p_0) \underbrace{\int_\mathcal{X} \log\{ \Gamma_{\CC,\RR}(x,p_0) \} f_{X\mid Y}(x\mid 0) dx}_{=\Psi_{\CC,\RR}(p_0,0)}  
=
\sC_{\CC,\RR}(p_0), 
\end{multline*} 
where $p_0$ is the only unidentified object on the right-hand side. Therefore, \cref{ass:rare} yields the identified bounds on $\bar\vartheta_\RR$ such that
\[
\min_{p\in[0,\bar p]} \sC_{\CC,\RR}(p) \leq \bar\vartheta_\RR \leq \max_{p\in[0,\bar p]}\sC_{\CC,\RR}(p),	
\]
where sharpness follows from the continuity of $\sC_{\CC,\RR}$.  The case of $\bar\vartheta_\AR$ is similar, and it will be omitted.  Now, consider the case of \cref{design2}. Then, we have $f_{X^*}(\cdot) = f_{X\mid Y}(\cdot\mid 0)$. Therefore, it follows from \cref{thm:unconfounding} that  
\[
\bar\vartheta_\RR = \int_\mathcal{X}  \log\{ \OR(x) \} f_{X\mid Y}(x\mid 0 ) dy = \Psi_{\CP,\RR}(0,0).	
\]
For $\bar\vartheta_\AR$, we note from \cref{thm:unconfounding} that
\[
	\bar\vartheta_\AR = \int_\mathcal{X} r_\CP(x,p_0) \Gamma_{\CP,\AR}(x,0) f_{X\mid Y}(x\mid 0) dx = \Psi_{\CP,\AR}(p_0),
\]
where $p_0$ is the only unidentified object on the utmost right-hand side. Finally, note that $\Psi_{\CP,\AR}(p)$ is a linear function in $p$ by the definition of $r_\CP(x,p)$: i.e., $\Psi_{\CP,\AR}(p) = C p$ for some constant $C$. Therefore, \cref{ass:rare} yields 
\[
\min\{ 0, C\bar p\} \leq \bar\vartheta_\AR \leq \max\{ 0, C \bar p	\}, 
\]
where sharpness follows from the continuity of $\Psi_{\CP,\AR}(p) = C p$.  \qed

\noindent\textbf{Proof of \cref{cor:aggregation2}: } 
Consider the case of \cref{design1}. From the proof of \cref{{thm:bounds}}, we know that for all $x\in \mathcal{X}$, 
\[
0 \leq \log\{ \theta_\RR(x) \} \leq \log\{ \Gamma_{\CC,\RR}(x,p_0) \},
\]
where the bounds are sharp if $p_0$ is given. Since $f_{X^*}(\cdot) = p_0 f_{X\mid Y}(\cdot\mid 1) + (1-p_0) f_{X\mid Y}(\cdot\mid 0)$ in this sampling scenario, we know that
\begin{multline*}
0
\leq 
\bar\vartheta_\RR 
\leq 
p_0 \underbrace{\int_\mathcal{X}\log\{ \Gamma_{\CC,\RR}(x,p_0) \} f_{X\mid Y}(x\mid 1) dx}_{=\Psi_{\CC,\RR}(p_0)} 	\\
+
(1-p_0) \underbrace{\int_\mathcal{X}\log\{ \Gamma_{\CC,\RR}(x,p_0) \} f_{X\mid Y}(x\mid 0) dx}_{=\Psi_{\CC,\RR}(p_0)}
=
\sC_{\CC,\RR}(p_0),
\end{multline*}
where $p_0$ is the only unidentified object on the utmost right-side expression. Therefore, the conclusion follows from \cref{ass:rare}, where sharpness follows from the continuity of $\sC_{\CC,\RR}$. The case of $\bar\vartheta_\AR$ is similar, and it will be omitted. Now, consider the case of \cref{design2}. Again, from the proof of \cref{thm:bounds}, we know that for all $x\in \mathcal{X}$, 
\[
0 \leq \log\{ \theta_\RR(x) \} \leq \log\{ \OR(x) \},
\]
where the bounds do not depend on $p_0$, but they are still sharp even if $p_0$ is given. Since $f_{X^*}(\cdot) = f_{X\mid Y}(\cdot\mid 0)$ in this sampling scenario, the sharp bounds on $\bar\vartheta_\RR$ will be given by 
\[
	0 \leq \bar\vartheta_\RR \leq \int_\mathcal{X} \log\{ \OR(x) \} f_{X\mid Y}(x\mid 0)dx = \Psi_{\CP,\RR}(0,0). 
\]
For $\bar\vartheta_\AR$, note from the proof of \cref{thm:bounds AR} that 
\[
0\leq \theta_\AR(x) \leq r_\CP(x,p_0) \Gamma_{\CP,\AR}(x,0),	
\]
where the bounds are sharp if $p_0$ is given.  Therefore, the sharp bounds on $\bar\vartheta_\AR$ with $p_0$ being given will be 
\[
	0\leq \bar\vartheta_\AR \leq \int_\mathcal{X} r_\CP(x,p_0) \Gamma_{\CP,\AR}(x,0) f_{X\mid Y}(x\mid 0) dx= \Psi_{\CP,\AR}(p_0). 
\]
Since $p_0$ is the only unidentified object, the conclusion follows from \cref{ass:rare}, where $\Psi_{\CP,\AR}(p) = C p$ for some $C\geq 0$, and sharpness follows from the continuity of $\Psi_{\CP,\AR}$.   \qed

\section{Remark on assumption \ref{ass:rare}}\label{appx:remark_pbar}
The $\bar p^*$ defined in \cref{ass:rare} satisfies that $0\leq \bar p^* \leq 1$. This can be seen from \begin{align*} 
1
&=
\sum_{t=0}^1 \int_{\mathbb{R}} f_{T,X|Y}(t,x\mid 0)\ dx
\geq
\sum_{t=0}^1\int_{\{(t,x):\ f_{T,X|Y}(t,x\mid 1) >0  \}} f_{T,X|Y}(t,x\mid 0)\ dx \\
&\geq
\sum_{t=0}^1 \int_{\{(t,x):\ f_{T,X|Y}(t,x\mid 1) >0  \}} \bar p^* f_{T,X|Y}(t,x\mid 1)\ dx 
=
\bar p^*
\geq
0,
\end{align*}
where the second inequality is because $\bar p^* \leq f_{T,X|Y}(t,x\mid 0)/f_{T,X|Y}(t,x\mid 1)$ for all $(t,x)$ with $f_{T,X|Y}(t,x\mid 1)>0$ by definition. We thank an anonymous referee for pointing this out to us.

%\setstretch{1.2}
%\bibliography{casecontrol_SL}
%\setstretch{\BASELINESTRETCH}

\end{document}